\documentclass[manuscript]{aastex}

\usepackage{upgreek}
\usepackage{graphicx}
\usepackage{natbib}
\usepackage{amsmath}
\usepackage[version=3]{mhchem} 
\usepackage{hyperref}
\usepackage{bm}

\newcommand{\upd}{\mathrm{d}}
\newcommand{\2}{$_2$}

\shorttitle{Condensible-rich convection}
\shortauthors{Feng et al.}

\begin{document}

\title{Convection in condensible-rich atmospheres}


\author{F. Ding}
\affil{Department of the Geophysical Sciences, University of Chicago, Chicago IL 60637, USA}
\email{fding@uchicago.edu}

\and

\author{R. T. Pierrehumbert}
\affil{Department of Physics, University of Oxford, Oxford OX1 3PU, UK }



\begin{abstract} 

Condensible substances are nearly ubiquitous in planetary atmospheres. For the
most familiar case -- water vapor in Earth's present climate -- the condensible
gas is dilute, in the sense that its concentration is everywhere small relative
to the noncondensible background gases. A wide variety of important planetary climate
problems involve nondilute condensible substances. These include planets near
or undergoing a water vapor runaway and planets near the outer edge of the conventional
habitable zone, for which $\mathrm{CO_2}$ is the the condensible. Standard
representations of convection in climate models rely on several approximations 
appropriate only to the dilute limit, while nondilute convection differs
in fundamental ways from dilute convection. In this paper, a simple parameterization
of convection valid in the nondilute as well as dilute limits is derived, and used
to discuss the basic character of nondilute convection. The energy conservation
properties of the scheme are discussed in detail, and are verified in radiative-convective
simulations. As a further illustration of the behavior of the scheme, results for
a runaway greenhouse atmosphere both for steady instellation and seasonally varying instellation
corresponding to a highly eccentric orbit are presented.
The latter case illustrates that the high thermal inertia associated with latent
heat in nondilute atmospheres can damp out the effects of even extreme seasonal forcing.

\end{abstract}

%
\keywords{convection --- planets and satellites: atmospheres --- planets and satellites: terrestrial planets}

%
%
%
\section{Introduction}\label{sec:intro}

Most planetary atmospheres contain one or more condensible substances,
which can undergo phase transitions from vapor to liquid or solid form
within the atmosphere.  The most familiar condensible is water vapor,
which plays a key role in Earth's climate, and in the runaway water
vapor greenhouse which is central to the past climate evolution of Venus
and also determines the inner edge of the habitable zone
\citep{ingersoll69,kasting88,nakajima92,kasting93,kopparapu13}. $\mathrm{CO_2}$ condensation is of importance
on both present and Early Mars, and plays a similar role to water vapor
in determining the outer edge of the habitable zone
\cite[Chapter~4]{ClimateBook}. Condensible substances are ubiquitous in planetary atmospheres;
additional examples include $\mathrm{CH_4}$ on Titan, $\mathrm{N_2}$ on Triton and possibly in Titan's past climates,  $\mathrm{NH_3}$ and
$\mathrm{NH_4SH}$ on Jupiter and Saturn, and $\mathrm{CH_4}$ on Uranus and Neptune. On hot Jupiters
and other strongly irradiated planets condensibles can even include a variety of substances more commonly
thought of as rocks and minerals in Earthlike conditions, such as enstatite ($\mathrm{MgSiO_3}$) or iron. 
In this paper, the term "moist" will be used to refer to any condensible substance,
not just water vapor. The diversity of climate phenomena associated with the choice of condensible
is augmented by the variety of relevant noncondensible background gases. For example, planets
with an $\mathrm{H_2}$ background gas are of considerable interest, and even Earth and Super-Earth sized 
planets can retain an  $\mathrm{H_2}$ atmosphere if they are in sufficiently distant orbits \citep{pierrehumbert2011hydrogen}.

In the present Earth's climate, water vapor is a dilute condensible, in
the sense that water vapor make up a small portion of any parcel of air.
For example, saturated air at a Tropical surface temperature of 300K
contains about 3.5\% water vapor, measured as a molar (also called
volumetric) concentration. The convection parameterizations in use in
conventional terrestrial general circulation models rely on a number of
approximations appropriate to the dilute limit.

\begin{figure}[h]
  \centering
  \includegraphics[width=0.7\textwidth]{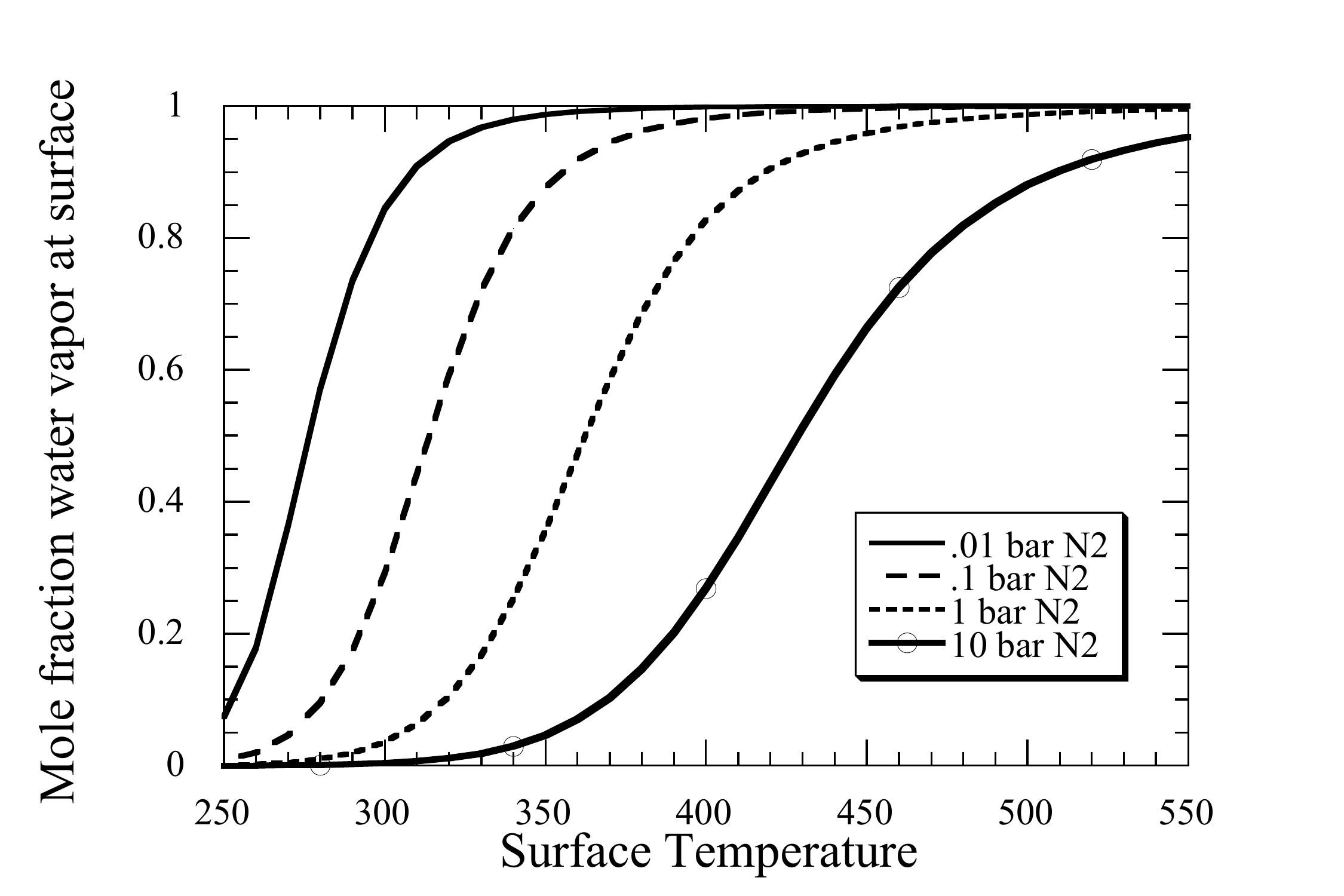}\\
	  \caption{The molar concentration of water vapor at the surface as a
  function of the surface temperature. The calculation assumes that the atmosphere is fully saturated following the
  moist adiabat. The calculation is done for an atmosphere consisting of water vapor and noncondensible
  $\mathrm{N_2}$, holding the total atmospheric mass of $\mathrm{N_2}$ fixed as surface temperature is changed. The amount of
  $\mathrm{N_2}$ in the atmosphere for each curve shown is given in terms of the surface pressure the $\mathrm{N_2}$ would have if it were
  present in the atmosphere alone (i.e. $g\cdot M_{N2}$, where $g$ is the acceleration of gravity and $M_{N2}$ is the mass of
  $\mathrm{N_2}$ per square meter of surface).  }\label{fig:mass_water}
\end{figure}

If the planet in question has a large condensed reservoir, such as an ocean or an icy
crust, then nondiluteness increases with temperature because, according to the
Clausius-Clapeyron relation, saturation vapor pressure for any substance increases approximately
exponentially with temperature.  Although a variety of dynamical and microphysical effects
prevent real atmospheres from attaining saturation, the actual condensible content nonetheless
tends to scale with saturation vapor pressure  \citep{pierrehumbert07}.  The nondiluteness
for a given temperature also depends on the mass of noncondensible background gas
in the atmosphere, since a massive atmosphere can dilute a greater quantity of condensible.
The way these two factors play out for the case of condensible water vapor in noncondensible
$\mathrm{N_2}$ is illustrated in Fig. \ref{fig:mass_water}. In the nondilute case, one must
take care with the way the quantity of noncondensible gas in the atmosphere is specified,
as the noncondensible partial pressure becomes dependent on altitude and -- for a fixed mass of
noncondensible -- varies with temperature as the amount of condensible gas in the atmosphere
changes. In this graph, we measure the amount of $\mathrm{N_2}$ in terms of the surface pressure the
$\mathrm{N_2}$ would exert if it were present in the atmosphere alone, and carry out the calculation
in such a way that this quantity (and hence the mass of noncondensible) remains fixed as temperature
changes. We call this quantity the $\mathrm{N_2}$ {\it inventory}, and refer to it by the symbol
$p_{\mathrm{N_2},I}$ (with analogous notation for other gases). For a given noncondensible inventory,
the temperature at which the atmosphere becomes nondilute is independent of the planet's surface gravity.
The corresponding mass of noncondensible gas A per square meter of the planet's surface, $p_{\mathrm{A},I}/g$,
is higher for planets with lower gravity, though.

The condensible fraction is greatest at the ground, so the atmosphere first becomes nondilute
there. Hence in Fig.~\ref{fig:mass_water} we show the ground level molar concentration of condensible. 
When the $\mathrm{N_2}$ inventory is 0.01 bar, the low level atmosphere is already roughly half
water at temperatures of 270K, whereas with a 1 bar  $\mathrm{N_2}$ inventory (somewhat greater than the present
Earth's) this concentration isn't attained until the surface temperature exceeds 350K.  With a 10 bar
inventory, the atmosphere is still dilute at 350K, and doesn't become strongly nondilute until
425K. Thus, the noncondensible inventory of an atmosphere is a crucial factor in determining the
character of moist convection.  This result underscores the importance of the factors governing
the noncondensible inventory on a planet.  \citet{wordsworth13} found that a high noncondensible
inventory can also inhibit loss of water, and the reasoning in that paper applies to other condensible
substances as well.

Temperature decreases with height on the adiabat, approaching zero as the pressure approaches zero, so within the
convective region of an atmosphere -- the troposphere -- the atmosphere will become more dilute with height even if it
is highly nondilute near the surface.  Even in the radiative-equilibrium portion of the atmosphere -- loosely speaking, the 
stratosphere -- in the absence of atmospheric absorption of incoming stellar radiation temperature decreases with altitude where the atmosphere is optically thick in some portion of the infrared spectrum. However, the temperature decrease
ceases where the atmosphere becomes optically thin in the infrared, and suitably strong absorption of incoming stellar
radiation can cause the temperature to increase with height, as in Earth's stratosphere \cite[Ch.~4]{ClimateBook}. 
Either effect can keep a nondilute lower atmosphere from being capped by a dilute upper troposphere or dilute stratosphere.
The factors governing the diluteness of the stratosphere were discussed in detail in \cite{wordsworth13}, 
in connection with loss of volatiles to space. 

All of the issues discussed in this paper are independent of the choice of condensible
and noncondensible gas, though the choice of gases does affect the surface temperature
at which the atmosphere begins to become nondilute, for any given inventory of
noncondensible gas.  For example, given a 1 bar inventory of noncondensible
gas, $\mathrm{CO_2}$ condensing in an $\mathrm{N_2}$ atmosphere reaches 50\% concentration
at the ground for a surface temperature of 195K, and $\mathrm{CH_4}$ condensing
in $\mathrm{N_2}$ reaches that concentration at 110K.  For $\mathrm{N_2}$ condensing
in an $\mathrm{H_2}$ background, the threshold is 95K.  
 
The most familiar case in which nondilute physics becomes important is in the water vapor 
runaway greenhouse.  This phenomenon has been extensively studied in one-dimensional
radiative-convective models, but there is an increasing need to explore three dimensional
simulations of this phenomenon.  Phenomena related to clouds, subsaturation and geographical 
temperature and constituent variations can only be adequately treated in the context of a
three dimensional general circulation model (GCM).  \cite{ishiwatari02} carried out pioneering
three-dimensional GCM studies of the runaway greenhouse using gray-gas radiation, and found that
subsaturation played an important role in determining the runaway threshold. 
More recently \cite{leconte13_nature} and \cite{yang2014strong}
carried out simulations incorporating clouds and real-gas radiation, finding that clouds as well as
subsaturation played a critical role. \cite{wolf2015evolution} carried out 3D simulations of a hot, moist Early Earth atmosphere
up to temperatures of 360K, which brings the lower troposphere well into the nondilute regime; convection in this case is
treated through a modification of the Zhang-Macfarlane convection scheme, but the nature of the modifications were not discussed.      None of these studies explicitly discussed the novel aspects of
convection in the nondilute case, or the physics involved in maintaining energy conservation. 
Generally speaking, there is a need for convection parameterizations
that cover the non-dilute limit accurately and are suitable for incorporation in GCM's.  There is also
a broad need for a better understanding of how moist convection operates in nondilute systems.

There are many other cases in which nondilute physics is important. Methane is a nondilute condensible
on Titan, making up about 30\% of its lower atmosphere when saturated.  $\mathrm{CO_2}$ is nondilute on Early Mars,
and more generally near the outer edge of the conventional habitable zone where habitability is supported
by the greenhouse effect of a thick $\mathrm{CO_2}$ atmosphere whose mass is limited by $\mathrm{CO_2}$ 
condensation onto the surface \citep{wordsworth581d-2011}. As is evident from Fig. \ref{fig:mass_water},
a low noncondensible inventory can cause water vapor to be nondilute even at temperatures comparable
to the present Earth; as such atmospheres are prone to losing water to space and generating abiotic oxygen \citep{wordsworth2014abiotic}, the dynamical features that could affect water loss are of considerable interest.

The dilute approximation enters into the formulation of convection
parameterizations used in conventional GCMs in several ways.  First, it
is used in the approximate form of moist enthalpy  to enforce energy
conservation in physical processes involving condensation.  Second, it
is invoked in order to neglect the energy carried away by precipitation.
 Third, it is used to justify the neglect of pressure changes that occur
when condensate mass is added to the atmosphere by evaporation or taken
away by precipitation . When the convection is nondilute, the pressure
effect can be important in driving atmospheric circulations. Moreover
removing mass from a layer of the atmosphere by precipitation unburdens
the layers below, causing them to expand and do pressure work.  
Finally, in the nondilute case there is a need to carefully take into
account the effect of changing atmospheric composition on mean
thermodynamic quantities such as specific heats or gas constants (though
some existing parameterizations already take such things into account,
at least approximately). 

Many of the novel aspects of nondilute convection affect energy conservation properties.
In situations where it is necessary to accurately track transient behavior (e.g. in seasonal
or diurnal cycles) or in which energy is transported from one geographical location
to another (as in a GCM), energy conservation becomes crucial. Even in a steady state
in a single-column model, energy conservation is important in models that explicitly
resolve the surface energy balance, since energy deposited at the surface is transferred
to the lowest model layer, whereafter it needs to be communicated upward by convection without loss or gain.
The only situation in which energy conservation in the course of convective adjustment
is unimportant is that in which detailed modeling of energy transfer between surface and
atmosphere is replaced by an assumption that the surface temperature equals the immediately overlying
air temperature, provided that only the steady state with infrared radiation to space balancing net absorbed
stellar radiation is of interest. This is a fairly common approach in radiative-convective models, which
has allowed many studies \citep[e.g.][]{kasting91} to avoid the necessity of confronting
energy conservation in nondilute convection. 

In this paper, we formulate and test a simplified convection parameterization scheme
that can deal with the full range of situations from dilute to condensible-dominated 
behavior, and use the scheme to explore some of the novel aspects of nondilute
moist convection. The scheme is suitable for use both in single-column radiative
convective models and in general circulation models. The scheme applies to any
combination of a single condensible substance and noncondensible background gas,
and conserves energy for any concentration of the condensible component. 

The development of our scheme builds in part on the treatment of convection in a pure $\mathrm{CO_2}$
Martian atmosphere given in \citet{forget98}, and the related scheme for thicker atmospheres 
employed in \citet{wordsworth581d-2011}.  \citet{leconte13_nature} incorporated some 
aspects of nondilute behavior in their 3D runaway greenhouse study, but did not specifically discuss the
convection scheme, its energy conservation properties or the nature of nondilute convection. 
Further development of the subject requires a broader discussion of the basic ways in which
nondilute convection differs from the familiar dilute case. 

In Section~\ref{sec:description} we present a complete description of
the convection scheme. We show that this scheme conserves energy whether
the condensible substance is dilute in the atmosphere or not.  As an
example of the behavior of the scheme, we apply it to an atmosphere
undergoing a runaway greenhouse on a planet with a high eccentricity
orbit in Section~\ref{sec:result}.  This serves as a test of energy
conservation in a situation involving strong transient forcing, where
key differences between dilute and nondilute convection, as revealed by
our simulations,  in Section~\ref{sec:discussion}, and summarize our
main findings, together with pointers toward unresolved issues,
in Section~\ref{sec:conclusions}.

\section{Description of the moist convection scheme}\label{sec:description}

\citet{manabe_1964} first introduced a convective adjustment scheme  in
their study of a single air column subject to atmospheric radiation and
convection. In their convection scheme, once the vertical lapse rate in
the model exceeds a critical value (e.g., 6.5\,K\,km$^{-1}$, a typical
value for present Earth's mid-latitude atmosphere), it is reset to the
critical value. This assumes  that the free convection is strong enough
to maintain the assumed critical lapse rate. The critical lapse rate
determines only the local slope of the adjusted $T(p)$ and an additional
principle must be invoked to determine the intercept. In the initial
version of the Manabe scheme, the column-integrated dry enthalpy
$\int_{0}^{p_s} (c_p T)\upd p/g $ is assumed to be conserved during the
convective adjustment,where $p_s$ is surface pressure, $c_p$ is the
specific heat of dry air and $g$ is the acceleration of gravity. Energy
stored in the form of latent heat was not taken into account.
\citet{manabe65} modified this convection scheme for a moist atmosphere
when studying the climatology of a GCM with a simple hydrological cycle.
This parameterization used the moist adiabat as the critical lapse rate
and conserved the column-integrated moist enthalpy in the dilute limit 
$\int_{0}^{p_s} (c_p T + L q)\upd p/g $ during the convective
adjustment, where $L$ is the latent heat of condensation and $q$ is the
mass mixing ratio of water vapor. \citet{betts_86} and
\citet{betts_miller_86} proposed another new convection scheme
(now known as the Betts-Miller scheme) by relaxing the temperature and humidity profiles
gradually towards the post-convective equilibrium states with a given
relaxation timescale. The precipitation rate and the change of the
temperature and humidity profiles during the convective adjustment are
again computed using conservation of the column-integrated moist
enthalpy in the dilute limit. These schemes are  widely used in
idealized experiments
\citep{renno94,frierson07,ogorman_hydrological_2008,merlis10}.

In this paper, we propose a new simplified convection scheme similar
in spirit to the convective adjustment scheme in  \citet{manabe_1964}
and the Betts-Miller scheme, but without assuming the condensible
substance to be dilute. The resulting scheme conserves non-dilute moist
enthalpy,  takes  the enthalpy of condensate and the mass loss from
precipitation into account, and works for both dilute and non-dilute
cases no matter what the condensable substance is.   

The atmosphere is divided into discrete layers, and the scheme is applied iteratively to
adjacent pairs of layers. The adjustment is performed for each pair of layers from bottom
to top, with the procedure being repeated until the 
temperature and moisture profiles converge to their asymptotic form within the specifed
degree of accuracy. 

When a pair of adjacent layers is found to be unstable to convection,
adjustment to a neutral state is performed through a two step process.
The subdivision of the adjustment into steps makes it easier to assure
that each step (and hence the whole process) conserves energy.  First,
the pair is adjusted to the reference neutrally stable state of
temperature and humidity conserving the summed nondilute moist enthalpy
and moisture of the two layers, assuming that the final state is
saturated in both layers. 
Condensate formed in this step is retained in
each layer in such a way that the total mass of each layer remains unchanged;
hence in this step the pressure of the interface between the layers and
of lower layers remains unchanged, in accordance with hydrostatic
equilibrium.  If there is not enough moisture present to saturate both
layers, the adjustment is done without forming any condensate, in a manner
that will be detailed later.  

In the second step, condensate is removed from the layers. For simplicity,
we currently assume infinite precipitation efficiency, so that all precipitation
is transported to the surface without any evaporation along the way; we also neglect
heating due to frictional dissipation associated with the falling precipitation. The
scheme can be easily modified to incorporate these complicating effects, but here
we wish to focus on the most basic effects of nondiluteness. When condensate is
removed, the atmosphere is allowed to adjust to the resulting change in pressure,
and the energy carried by precipitation is tracked, in such a way that the
system conserves energy in the course of precipitation. 

When mass is removed from the atmosphere by precipitation, the potential
and thermal energy of the precipitation must be added to the surface energy
reservoir for energetic consistency. At the bottom boundary, surface fluxes
remove energy and condensate mass from the surface reservoir, and add it to
the lowest model layer. In Sections \ref{subsec:mass} and \ref{subsec:evap}
we show how the energy budget is balanced in the course
of these processes.

\subsection{Thermodynamic preliminaries: Nondilute moist enthalpy and the nondilute pseudoadiabat}

We begin with a review of some relations pertinent to energy conservation in the nondilute
case.  The calculations involve three atmospheric constituents:  a noncondensible gas (or mixture
of gases) referred to with subscript $a$, and a condensible substance whose gaseous form
is referred to by subscript $c$ and whose condensed form is referred to by subscript $\ell$. 
The subscript $t$ will be used to refer to quantities associated with the total mass of condensible
in both phases. 

The temperature dependence of latent heat enters the problem at several points. 

When the density of the condensate is much greater than that of the
vapor phase, the specific latent heat varies linearly with the
temperature \citep[][p. 115]{emanuel94}. 
\begin{equation}\label{eqn:LvsT}
\frac{\upd L}{\upd T} \approx c_{pc} - c_{p\ell}
\end{equation}
Here subscript $\ell$ represents the condensate, which may be in a solid or liquid phase. We will take this temperature dependence
into account in our formulation, but assume the specific heats to be independent of temperature. 

Next consider a parcel of atmosphere which has become supersaturated by one means or
another, and which relaxes back to saturation through condensation without loss of energy or
mass. Thus 
\begin{equation}
\upd m_c + \upd m_\ell = 0, \qquad \upd m_c < 0
\end{equation}
where $m_c$ and $m_\ell$ are the mass of condensible gas and condensate in the parcel. Suppose
further that the pressure is kept constant in the course of the condensation. In hydrostatic
equilibrium, this would correspond to the situation in which no mass is lost from the column of
atmosphere above the parcel in question.  The First Law of Thermodynamics then implies
\begin{equation} \label{eq:FirstLaw}
	0 = L(T) \upd m_c + (c_{pa}m_a + c_{pc}m_c + c_{p\ell}m_\ell)\upd T
\end{equation}
where $m_a$ is the mass of noncondensible in the parcel. 
From Eq. \ref{eqn:LvsT}  and the assumption that specific heats
 do not depend on temperature. we deduce
\begin{equation} \label{eq:HeatDifferential}
	L(T) \upd m_c = \upd [L(T)\  m_c] - (c_{pc} - c_{p\ell}) m_c \upd T 
\end{equation}
Combining this result with Eq. \ref{eq:FirstLaw} we can define a quantity which is conserved in the course of isobaric condensation without loss of mass or energy: 
\begin{equation} \label{eq:moistk}
\tilde{k} = (c_{pa}m_a + c_{p\ell}m_t)T + L(T)m_c,
\end{equation}
where $m_t = m_c + m_l$.
The conserved variable $\tilde{k}$ is usually referred as the moist enthalpy \citep[][p.~118]{emanuel94}.
Dividing by $m_a+m_t$ yields the moist enthalpy per unit total mass
\begin{equation} \label{eq:moistk}
k = (c_{pa}q_a + c_{p\ell}q_t)T + L(T)q_c,
\end{equation}
where $q$ is the mass concentration of the atmospheric constituent indicated by each subscript.
Allowing for pressure work done by the parcel in the general case, the First Law becomes
\begin{equation}\label{eqn:FirstLawP}
  \upd k - \frac{1}{\rho} \upd p = \delta Q
\end{equation}
where $\rho$ is the total density of all constituents and $\delta Q$ is the energy change due to such
processes as radiative heating or cooling. Alternately, in a temperature-volume formulation the
First Law can be written
\begin{equation}\label{eqn:FirstLawV}
  \upd \left(k - \frac{p}{\rho}\right) + p \upd\frac{1}{\rho} = \delta Q
\end{equation}

Eqns. \ref{eqn:FirstLawP} and \ref{eqn:FirstLawV} are conservation law that apply following an individual air parcel, but
that is not generally the same as the conservation law that applies when a column consisting of
various air parcels is mixed by convection.  Even if the initial and final states of the
column are in hydrostatic balance, convection proceeds through nonhydrostatic motions,
so energy conservation must be formulated to allow for nonhydrostatic dynamics. This is most
easily done using altitude ($z$) rather than pressure ($p$) as the vertical coordinate.  In
Appendix \ref{apx:conservation} it is shown that an isolated volume of atmosphere conserves the
quantity
\begin{equation}\label{eqn:ColumnEnthalpyZ}
\int_{z=0}^\infty \rho\cdot \left( k - \frac{p}{\rho} + gz \right) \upd z
\end{equation}
upon mixing by fluid motions of an arbitrary type, provided any condensate formed is retained within
the air parcel in which it forms (though it is free to evaporate back into the air parcel). The
density $\rho$ in this formula includes the mass of condensate. This proceeds
from a small variant of a standard thermodynamic derivation, but it is reproduced in the Appendix for the
sake of checking that it remains valid even in the nondilute limit, and even in the presence of condensate.
The conservation law given in Eq. \ref{eqn:ColumnEnthalpyZ} is an approximate form of the exact conservation
law, valid when kinetic energy in the initial and final states is negligible compared to the thermal energy.
The energy contains contributions both from the internal energy $k - {p}/{\rho}$ and
potential energy $gz$.  In the dilute case for an ideal gas, $q_a \approx 1$, ${p}/{\rho} \approx R_a T$,
$q_t \ll q_a$ and the internal energy takes on the familiar form $ c_{va} T + Lq_c$. Returning to
the general case, if the column is in hydrostatic balance, then $g\rho \upd z = -  \upd p$ and
\begin{equation}
\int_{z=0}^\infty \rho g z \upd z = \int_{p=0}^{p_s} z \upd p = -  \int_{p=0}^{p_s} p \upd z = \int_{z=0}^\infty p \upd z
\end{equation}
where $p_s$ is the surface pressure. Thus, for a column in hydrostatic balance the potential energy
term cancels the term in Eq. \ref{eqn:ColumnEnthalpyZ} arising from ${p}/{\rho}$ whence we conclude
that the quantity
\begin{equation}\label{eqn:ColumnEnthalpyP}
\int_{p=0}^{p_s}  k  \upd p
\end{equation}
is conserved in the course of mixing of a column of atmosphere provided the initial and final
states are in hydrostatic balance and no mass is lost from any air parcel in the course of the mixing.
The potential energy does not appear explicitly in this expression, but it is implicitly accounted
for through the fact that $c_p$ rather than $c_v$ appears in the expression for enthalpy $k$. This will
be important when we come to consider the energy carried by precipitation that removes mass from the
atmosphere. One knows intuitively that part of the energy carried by the precipitation should be
in the form of potential energy -- that is after all where hydroelectric power comes from, namely
the potential energy of water vapor stored in the atmospehre when solar energy is used to drive convection
which lifts the water vapor to higher altitudes. Without carefully considering the above derivation,
it is not clear where this potential energy is taken from when mass is removed from the atmosphere. 
The energy book-keeping is transparent if done in $z$-coordinates but is more subtle when done in
$p$ coordinates for a hydrostatic atmosphere. We will return to this point in Section \ref{subsec:mass}
where we deal with mass loss. 

We use the pseudoadiabat \citep[][p.~129]{ClimateBook} as the
reference temperature profile, meaning that the condensate does not
accumulate in the atmosphere to the extent that the temperature profile
would be significantly affected. By substituting the expression for
$\upd\ln p_a/\upd\ln p$ into the formula given in \citep[][p.~129]{ClimateBook}, the
pseudoadiabatic slope can be written in the form
\begin{equation} \label{eq:moist}
\frac{\upd \ln p}{\upd \ln T} = \frac{p_{sat}}{p} \frac{L(T)}{R_c T} + \frac{p_a}{p} \frac{c_{pa}}{R_a} \frac{1 +\left( \frac{c_{pc}}{c_{pa}} + \left( \frac{L}{R_c T} -1\right)\frac{L}{c_{pa} T} \right)r_{sat}}{1 + \frac{L}{R_a T} r_{sat}}
\end{equation}
This expression remains valid even in the nondilute case. 
Here $p_{sat}$ is the saturation vapor pressure, $L$ is the specific
latent heat of vaporization, $R$ is the specific gas constant, $c_p$ is the specific heat
capacity at constant pressure, $r_{sat}$ is the saturation mass mixing ratio.
Note that the pseudoadiabat
reduces to the Clausius-Clapeyron relation at temperatures high enough
that $r_{sat} \gg 1$, in which case $p\rightarrow p_{sat}$ and
${p_a}/{p} \rightarrow 0$.

\subsection{Lapse rate adjustment with retained condensate} \label{subsec:lapse}

\begin{figure}[h]
  \centering
  \includegraphics[width=0.7\textwidth]{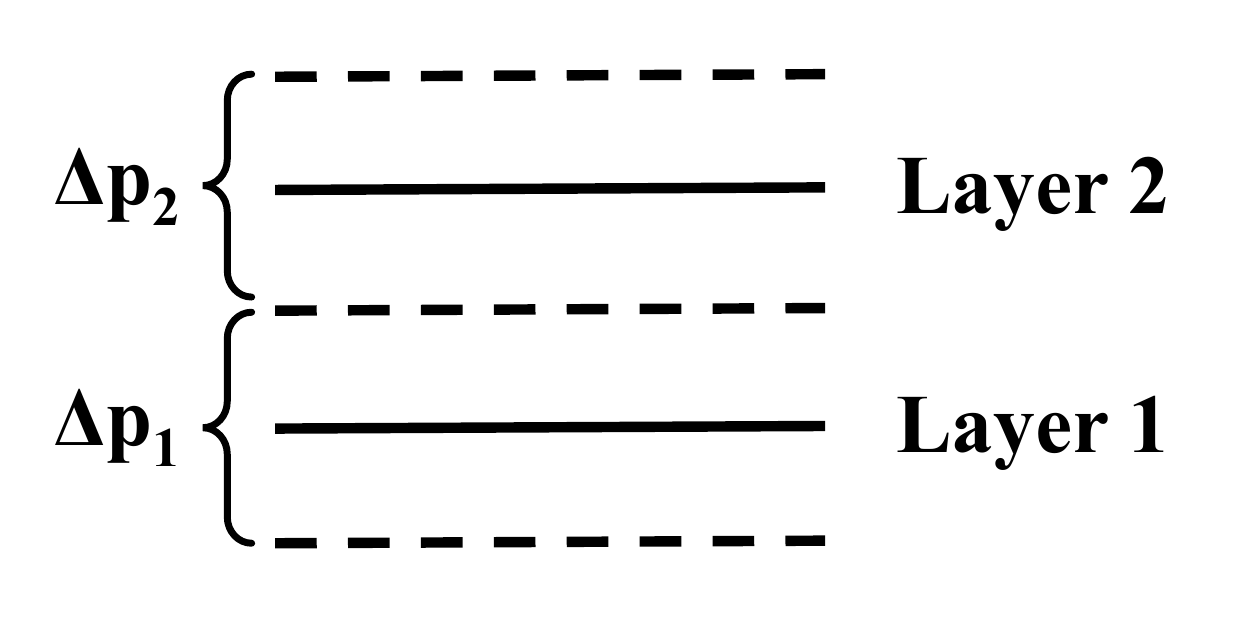}\\
  \caption{Notations used for the moist convective adjustment.}\label{fig:layers}
\end{figure}

Consider two adjacent layers of atmosphere arranged as in Fig. \ref{fig:layers}. The
lower layer has pressure and temperature $(p_1,T_1)$ and pressure thickness $\Delta p_1$,
while the upper layer has pressure and temperature $(p_2,T_2)$ and pressure thickness $\Delta p_2$.

In the first step of the convection scheme, we check the temperature difference between 
the two layers. If it is steeper than the pseudoadiabat computed on the basis of 
Eq. \ref{eq:moist}, heat and moisture is assumed
to mix between the two layers, resulting in a state that satisfies the following criteria:
\begin{itemize}
    \item The temperature difference between the layers is given by the slope of
          the pseudoadiabat.
    \item The vapor phase of the condensible is saturated in each layer,
          provided that the mass of condensible in the adjusted state is
          less than or equal to the available condensible mass in the initial state.
    \item The excess of condensible mass between the initial and final state
          is turned into condensate, which is retained within the two layers.
    \item The net nondilute moist enthalpy of the two layers is the same as in the initial state.
    \item The net mass in each layer remains unchanged, so according to the
          hydrostatic relation,  $\Delta p_1$ and $\Delta p_2$ are
          the same in the initial and adjusted states. 
\end{itemize}
The last of these assumptions is not really a physical assumption, but rather a convention about
how the pressure layer interfaces are labeled after mixing takes place. 
This set of requirements gives rise to a nonlinear relation with one
unknown parameter,which can be taken to be the temperature of the lower layer
after adjustment; this parameter is adjusted using a Newton iteration
until the conditions are satisfied to within a specified accuracy. 

Specifically let $T_{1,I}$ be the lower layer temperature for the initial state and
$T_{1,F}$ be the lower layer temperature for the adjusted state, with analogous notation
for the other quantities.  The temperature difference $\Delta T \equiv T_1-T_2$ on the pseudoadiabat
is computed from Eq. \ref{eq:moist}, and if $T_{1,I} - T_{2,I} > \Delta T$ convection is initiated
and the temperatures are reset to 
\begin{equation}
T_1 = T_{1,F}, T_2 = T_{1,F} - \Delta T
\end{equation}
where $T_{1,F}$ is yet to be determined. 
Because condensate forms in the course of the adjustment, it is most convenient to
express the saturation assumption in terms of the mixing ratios $r_c$ of gas-phase 
condensible relative to the noncondensible component. 
\begin{equation}
r_{c1,F} = r_\mathrm{sat}(T_{1,F},p_1), r_{c2,F} = r_\mathrm{sat}(T_{2,F},p_2)
\end{equation}
where $r_\mathrm{sat}(T,p)$ is the function giving the mixing ratio at saturation.
Conservation of mass of condensible substance between the initial and adjusted states requires
\begin{equation}\label{eqn:CondMassBalance}
\frac{r_{c1,F} + r_{\ell 1,F}}{1+r_{c1,F} + r_{\ell 1,F}} \Delta p_1 + \frac{r_{c2,F} + r_{\ell 2,F}}{1+r_{c2,F} + r_{\ell 2,F}} \Delta p_2 = q_{c1,I}\Delta p_1 + q_{c2,I} \Delta p_2
\end{equation}  
since there is assumed to be no retained condensate in the initial state. This yields only one constraint
for the two unknowns $r_{\ell 1,F}$ and $r_{\ell 2,F}$.  Because total mass is conserved in each layer,
the expression for conservation of noncondensible mass does not yield an independent relation. In order
to close the problem, an assumption must be made regarding the distribution of condensate between
the two layers.  Defining $\eta \equiv r_{\ell 2,F}/r_{\ell 1,F}$ allows Eq. \ref{eqn:CondMassBalance} 
to be solved for $r_{\ell 1}$, whence multiplication by $\eta$ gives $r_{\ell 2}$. The left hand
side of Eq. \ref{eqn:CondMassBalance} is monotonic in $r_{\ell 1}$ and approaches
$\Delta p_1 + \Delta p_2$ for $r_{\ell 1} \rightarrow \infty$, which is guaranteed to be greater
than the right hand side because $q_c \le 1$.  If the initial state has enough condensible to
saturate both layers, then the left hand side is less than the right hand side when 
$r_{\ell 1} = r_{\ell 2} = 0$, whence we conclude there is a unique solution with positive
$r_{\ell}$.  If this condition
is not satisfied, there is no physical solution. We will discuss the handling of that case shortly.

With the above assumptions, it is possible to compute the enthalpies $k_{1,F}$ and $k_{2,F}$ in
the adjusted layers given $T_{1,F}$.  Then $T_{1,F}$ is adjusted using a Newton iteration
until the enthalpy conservation relation
\begin{equation}
k_{1,F}\Delta p_1 + k_{2,F} \Delta p_2 = k_{1,I}\Delta p_1 + k_{2,I} \Delta p_2
\end{equation}
is satisfied.

There is no clear physical basis for determining $\eta$ but we do not believe it is a critical
parameter of the scheme, since all condensate is removed in the second stage of the adjustment process. 
In the calculations reported in this paper we adopted the choice
\begin{equation}
\eta = \frac{1+r_{c2}}{1+r_{c1}}
\end{equation}
which implies that condensate mass is distributed in proportion to the gas-phase mass of the atmosphere.
This was found to yield a stable iteration, and also has the virtue of allowing Eq. \ref{eqn:CondMassBalance}
to be solved analytically for $r_{\ell 1}$ in terms of a simple linear equation.

It sometimes happens that there is not enough moisture in the initial state to allow a saturated
adjusted state to be realized, i.e. that the adjusted saturated state would require negative
precipitation.  In such a case we still adjust the temperature to the pseudoadiabat in an
enthalpy conserving way, but do not form any precipitation. In this case, the condensible vapor is redistributed
in proportion to the final state saturation mixing ratio, with a proportionality constant $f_1 < 1$ 
chosen to conserve the condensible mass between the initial and final state:
\begin{equation}
r_{c1,F} = f_1 r_\mathrm{sat}(T_{1,F},p_1), r_{c2,F} = f_1 r_\mathrm{sat}(T_{2,F},p_2)
\end{equation} 
The temperature and moisture adjustment employed here 
is similar to the treatment of shallow nonprecipitating convection in 
the Simplified Betts Miller scheme \cite{frierson07} used in many conventional general circulation model
studies. 

Condensation can also happen in the absence of convection, and such condensation is generally 
referred to as large-scale condensation.  In a 3D general circulation model it can happen 
as a result of uplift and adiabatic cooling caused by the resolved large scale circulation,
but it can also be caused by radiative cooling; the latter mechanism is the only large scale
condensation mechanism in a 1D radiative-convective model.  When large scale condensation forms, it
can simply be added to the condensate (if any) produced by the convection step, and dealt with
in the precipitation scheme described in the next section.  

\subsection{Precipitation and mass loss from the atmosphere} \label{subsec:mass}

In the precipitation step, condensate mass is removed from the atmosphere one layer at
a time. The precipitation is assumed to reach the surface instantaneously without 
loss of energy or mass along the way. When this happens, an amount of enthalpy
$c_{p,\ell} T q_\ell\Delta p/g$ is removed from the layer and added to the surface
energy budget. The temperature of the remaining gas in the layer remains unchanged,
but the layer thickness $\Delta p$ goes down in accord with the mass loss from the layer.
This reduces the pressure of all the layers located at lower altitudes.  When condensate
is removed, it is like taking a brick off the top of a piston supporting a column of air;
the piston rises, and the air in the cylinder below the piston expands adiabatically until
the reduction in pressure force equals the new force of gravity exerted by the mass of the
piston. In the course of the expansion, pressure work is done. 

In the context of an atmospheric column, the enthalpy of layers below the one from which condensate
was removed goes down to compensate for the pressure work done. This reduction is manifest as
a reduction in the temperature of the lower layers, but because no mass is taken away from the
lower layers in this step, their individual pressure thicknesses remain unchanged. The temperature
reduction typically leads to supersaturation, but that is dealt with at the next condensation and
convection step. 

The pressure work done in the course of the expansion of the lower layers, and hence the
enthalpy reduction in those layers, can be computed as follows. Let $p_1$ be the pressure of
the level from which a condensate mass (per unit area) $\delta m$ was removed, and let $z_1$ be
the corresponding altitude.  Then the work per unit mass done at some level $p<p_1$ is
\begin{equation}
\frac{1}{\rho} \delta p = \frac{1}{\rho} g\delta m
\end{equation}
Then, doing a mass-weighted integral of this over all the lower layers yields the pressure work
\begin{equation}
\delta W \equiv \int_{p_1}^{p_s} \left[\frac{1}{\rho} \delta p\right] \frac{1}{g} \upd p 
 =  g\delta m \int_{p_1}^{p_s} \frac{1}{\rho}\frac{1}{g} \frac{\upd p}{\upd z} \upd z 
= g\delta m \int_0^{z_1} \upd z = g z_1 \delta m  
\label{eqn:PrecipPE}
\end{equation}
This is precisely the potential energy of the precipitation removed. Hence, energy conservation in
the column is achieved if the potential energy, as well as the enthalpy, of the precipitation is 
added to the surface budget.  In our simplified model, the potential energy is implicitly converted
to kinetic energy of falling precipitation, which is then dissipated as heat when the precipitation
strikes the surface.  \citet{forget98} also included the potential energy of precipitation in
the convection scheme used in their study of $\mathrm{CO_2}$ snowfall on Mars, but did not 
explicitly consider the way the inclusion of this term leads to energy conservation.

\subsection{Treatment of evaporation}\label{subsec:evap}

Evaporation is most easily treated in terms of discrete layers, rather than trying to pass to the continuum limit.  Evaporation from the surface reservoir of condensible adds condensible mass to the lowest layer
of the atmosphere, increasing its gas-phase condensible content and also increasing its mass, and hence
the pressure thickness $\Delta p$ of the layer. The pressure of higher layers is not affected, though implicitly
potential energy is being stored in higher layers since, viewed in $z$ coordinates, the addition of mass to 
the lowest layer raises the altitude of all overlying layers, insofar as the added mass takes up space. 
Additional energy is stored in the lowest model layer in the form of the latent heat of the vapor added to the layer,
which is equal to the energy lost from the planet's surface through evaporation.
 
Since the temperature dependence of the latent heat is taken into account, it may seem that the latent heat change during atmospheric condensation aloft is larger than the latent heat put into the atmosphere during surface evaporation of the same amount of mass. However, the thermal energy change on converting a mass $\delta m_\ell$ from vapor to liquid at temperature
$T$ is
\begin{equation}
\left[ L(T) + (c_{pl} - c_{pc})T(p) \right] \delta m_\ell
\end{equation}  
Which reflects th loss of sensible heat from the vapor phase and the gain in the form of sensible heat of the condensate,
as well as the latent heat of phase change.  From Eq. \ref{eqn:LvsT} for the temperature dependence of the latent heat,
the term in brackets is simply $L(0)$, which is independent of temperature. Hence, the {\it net} energy change from
condensation or evaporation is independent of temperature; the difference between latent heat added near the surface and
latent heat released aloft is accounted for in the sensible heat carried away by the precipitation.

\section{Results} \label{sec:result}

We carry out three experiments using a 1D column model incorporating the convection scheme introduced in Section~\ref{sec:description}. The first two simulations verify that the convection scheme conserves energy and mass for both dilute and non-dilute cases. In the third simulation we apply the model to the atmosphere undergoing the runaway greenhouse on a planet with a high eccentricity orbit, and illustrates some additional features of convection in the nondilute limit.  

\subsection{Model framework}

The non-condensable substance in the model is the mixed N\2-O\2 air on Earth, and the mass of the non-condensable substance, unless otherwise noted, is 10$^4\,\mathrm{kg\,m^{-2}}$, as in the present Earth's atmosphere. The condensable substance in the model is water. The moist convection scheme determines the level of the tropopause. We then calculate the level of the cold trap where the saturation concentration of water vapor reaches its minimum above the tropopause and assume uniform  concentration above the cold trap and relative humidity of unity between the tropopause and the cold trap. This is equivalent to a simplified vertical diffusion scheme. The calculation is performed with a gravitational acceleration of 9.8 $\mathrm{m/s^2}$. 

The column model uses a gray radiation scheme similar to that described in \citet{merlis10} in the longwave spectral region.  As in \citet{merlis10}, the absorption coefficient of water vapor is kept constant at  $0.1\,\mathrm{m^2\,kg^{-1}}$, but unlike
\citet{merlis10} we compute the water vapor optical thickness based on the actual water vapor path in the model, rather 
than an idealized profile. With the stated absorption coefficient,  the effective radiating level 
is 0.98\,hPa for pure water vapor atmosphere. Therefore, in order to resolve the radiative and convective process at high water concentrations, we choose the level of 0.01\,hPa as the top of the column model. In addition to the water vapor opacity,
which varies with the water content (and hence temperature) of the atmosphere, the model includes a fixed background
opacity which own would give the atmosphere a longwave optical thickness  $\tau_0 = 1.2$. This can be thought of as the opacity
due to a noncondensible trace gas in the atmosphere, such as $\mathrm{CO_2}$. With these parameters the curve of the outgoing longwave radiation (OLR) as a function of the surface temperature exhibits
overshoot for a saturated atmosphere, peaking at a value of OLR$_{max} = 258\,\mathrm{W\,m^{-2}}$ at $T= 320\,\mathrm{K}$ before asymptoting
to a value OLR$_\infty = 210\,\mathrm{W\,m^{-2}}$ at high $T$ (see Supplementary Material).  Thus, if the absorbed stellar radiation is
below $210\,\mathrm{W\,m^{-2}}$ the system can never enter a runaway greenhouse, and if it is above $258\, \mathrm{W\,m^{-2}}$ the system runs away regardless of the initial condition. For absorbed stellar radiation between 210\,$\mathrm{W\,m^{-2}}$  and $258\,\mathrm{W\,m^{-2}}$ the system will runaway if initialized warmer than 320\,K but will be attracted to a metastable non-runaway state if initialized cooler
than 320\,K.    

We assume the atmosphere is transparent to the incoming stellar radiation.  In reality the near-infrared absorption and Rayleigh scattering by water vapor molecules is very important when the water content in the atmosphere is high, but here we are chiefly interested in illustrating the behavior of the convection scheme, particularly with regard to vertical energy and moisture transport.
The relevant phenomena are brought out most clearly in the idealized case in which incoming energy is deposited only at the surface.
 
The simulations shown are
carried out with 5 iterations per time step and a time step of 40 minutes in the dilute and runaway simulations, and 20 minutes
in the equilibrium nondilute simulation.  

The lower boundary of the column model is a heat reservoir with fixed thickness of 1\,m and specific heat equal to that of liquid water. This lower boundary is included as a computational device to allow the surface energy budget to relax smoothly towards equilibrium; while it is mathematically identical to the common representation of an ocean mixed layer, it is not intended to represent either a shallow ocean or the mixed layer of an actual deep ocean. To represent an actual ocean layer, one would need to allow the depth to change in response to the mass budget, which is a simple modification, but an unnecessary complication in view of the use to which the heat reservoir is put in our model.    At the surface, the sensible and latent heat fluxes are computed by standard drag laws assuming constant  drag coefficients and surface wind velocity. The surface is also heated or cooled by absorption of the shortwave radiation transmitted through the atmosphere, and the net 
surface infrared flux (computed within the gray-gas model). The chief novel aspect of the surface budget in the nondilute
case is that the enthalpy of the precipitation is added to the surface budget, as is the kinetic energy of the precipitation
(after being converted to heat). In addition, surface pressure is updated by the net evaporation rate at the surface at the end of each time step.

\subsection{Dilute simulation} \label{subsec:dilute}

In the dilute simulation, we use the average shortwave solar flux absorbed by the Earth's climate system  $S_0 = 238\,\mathrm{W\,m^{-2}}$ as the insolation and assume the albedo of the surface is zero. With the assumed background
noncondensible opacity the equilibrium surface temperature in this simulation is similar to Earth's present tropics,
and the system is therefore in the dilute regime. 

\begin{figure}[h]
  \centering
  \includegraphics[width=\textwidth]{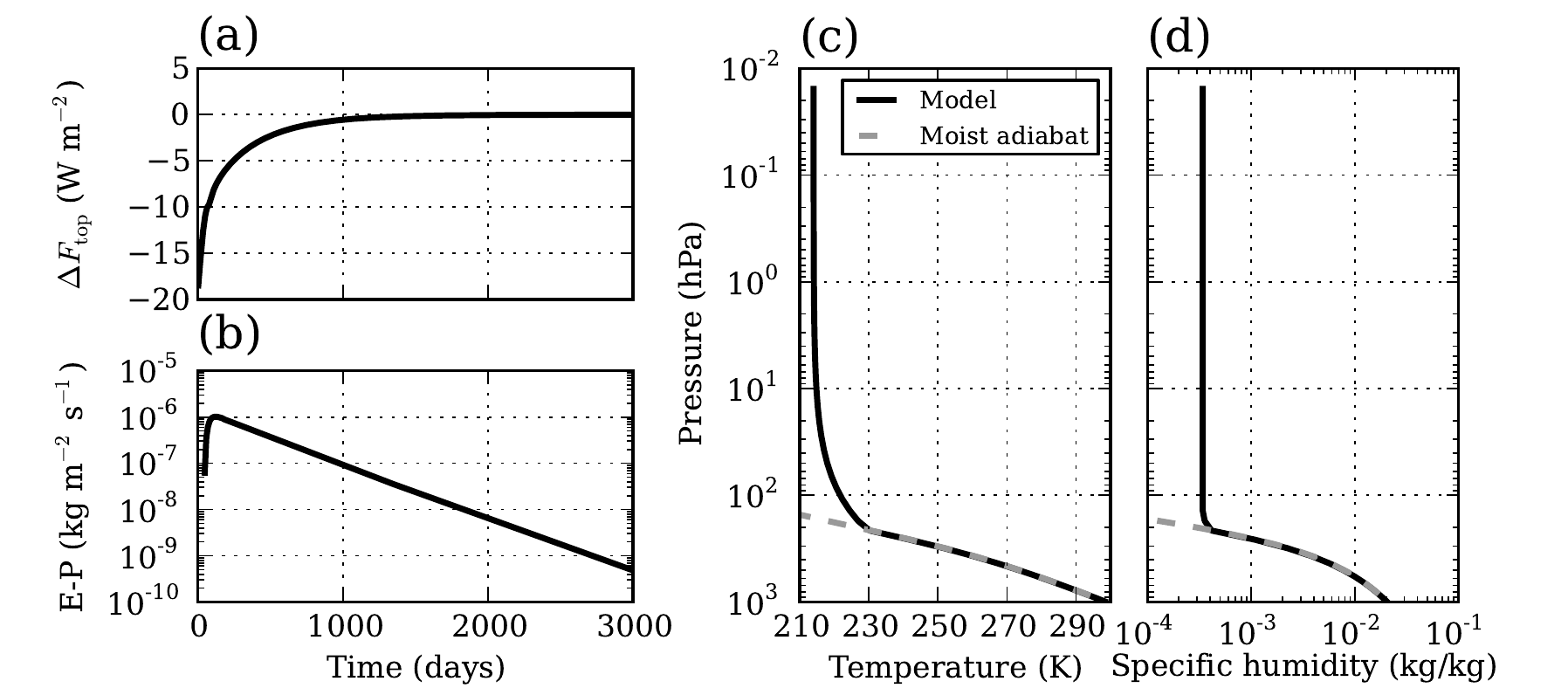}\\
  \caption{Time evolution of the net radiative flux at the top of the model (a) and the net evaporation flux at the surface (b) when insolation $S_0 = 238\,\mathrm{W\,m^{-2}}$, and the vertical profile of temperature (c) and specific humidity (d) when the column model reaches equilibrium. The dashed line in panel (c) gives the temperature on the moist adiabat and the dashed line in (d) gives the corresponding moisture in saturation.}\label{fig:time_dilute}
\end{figure}

We first check whether the convection scheme conserves energy.  Both the  net radiative flux at the top of the model and the net evaporation flux at the surface nearly vanish after integration of 3000\,days, implying that the system reaches not only energy but also mass equilibrium (Figure~\ref{fig:time_dilute}a and \ref{fig:time_dilute}b). 
The equilibrium profiles of air temperature and specific humidity are shown in Figure~\ref{fig:time_dilute}c and \ref{fig:time_dilute}d.  
Below $\sim$200\,hPa the atmosphere follows the moist adiabat, which defines the convective region. Above the tropopause, the atmosphere is in radiative equilibrium. Since there is no ozone in the column model, the upper atmosphere is nearly isothermal due to the gray gas assumption. The specific humidity at the surface is $\sim 2.08\times10^{-2}$\,kg/kg, confirming that water vapor is dilute in this simulation.

\begin{figure}[h]
  \centering
  \includegraphics[width=\textwidth]{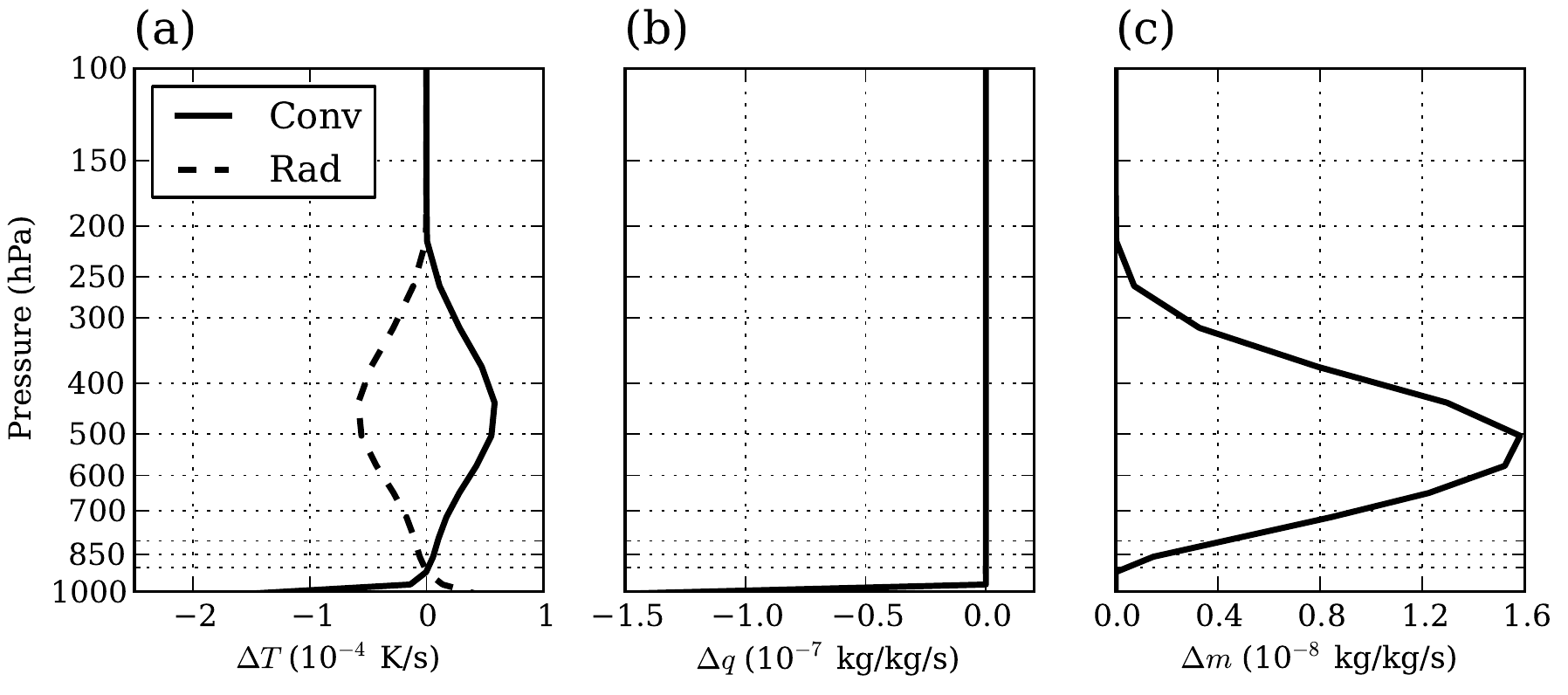}\\
  \caption{
  The rate of change of air temperature (solid line in a) and specific humidity (b), and the condensation rate of water (c) during the convective adjustment when the column model reaches equilibrium for $S_0 = 238\,\mathrm{W\,m^{-2}}$. The dashed line in (a) shows the radiative cooling profile. Near the surface, both the air temperature and specific humidity decrease during the convective adjustment primarily due to the upward transport of sensible heat and moisture. }\label{fig:convec_dilute}
\end{figure}

The behavior of the convection scheme is shown in Figure~\ref{fig:convec_dilute}. The troposphere is heated by the latent heat release and the convergence of sensible heat flux,  except near the surface. In equilibrium, this heating is balanced by the radiative cooling in the troposphere. 
Near the surface,   most of the sensible heat is transported upward and  little condensation occurs, resulting in a cooling effect
which is balanced by heat input from the surface (which is in turn heated by solar absorption).  The maximal convective warming is located at $\sim$500\,hPa, slightly higher than where most of the condensate forms. 

In this 1D column model, the water vapor profile is only updated in the surface evaporation and convection scheme. Therefore, in the steady state, the water vapor concentration does not change during the convective adjustment, except at the lowest model layer. In this layer, convective transport of moisture upward dries the layer (Figure~\ref{fig:convec_dilute}b), which is balanced by moisture input from evaporation.   An equivalent amount of water vapor  then condenses out as air parcels rise in the atmosphere (Figure~\ref{fig:convec_dilute}c), and falls to the ground, closing the mass budget of the climate system.

\subsection{Non-dilute simulation} \label{subsec:non}

\begin{figure}[h]
  \centering
  \includegraphics[width=\textwidth]{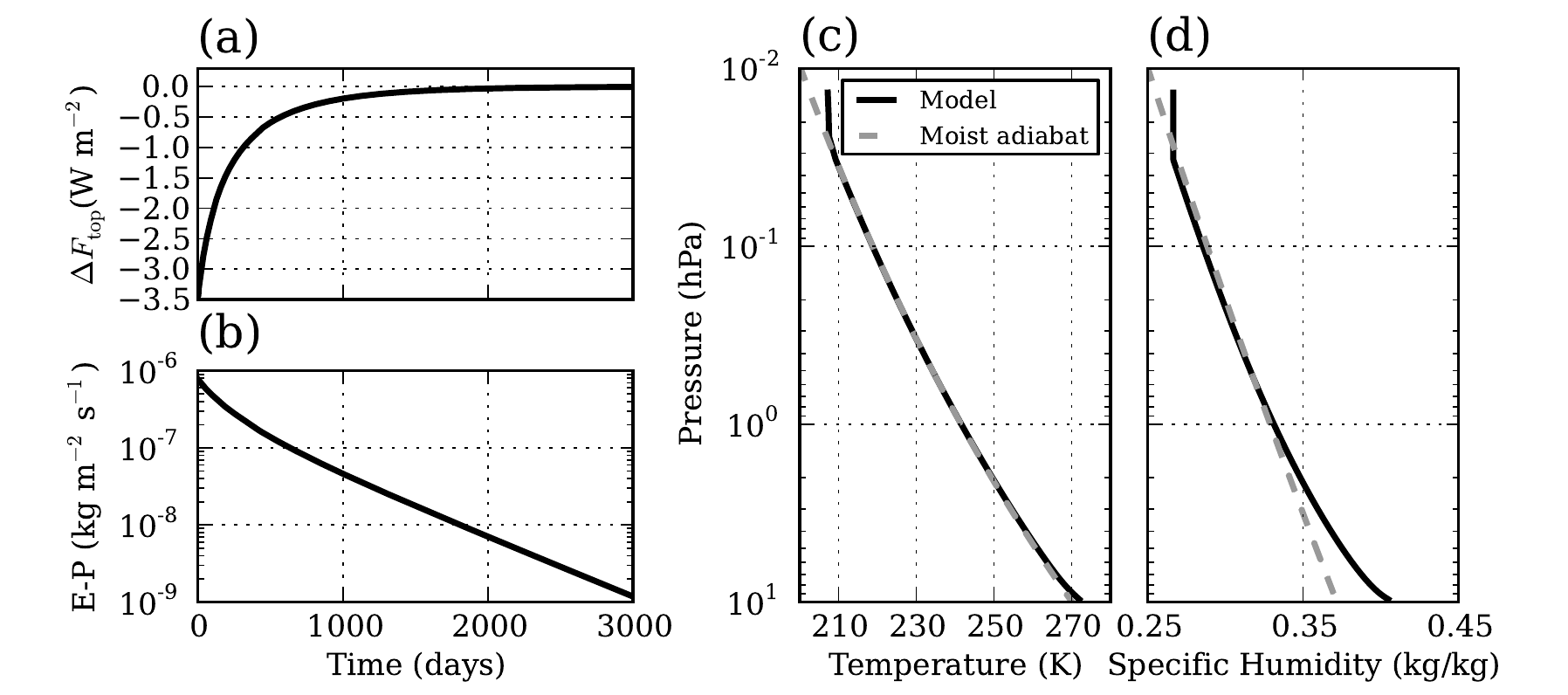}\\
  \caption{Same as Figure~\ref{fig:time_dilute}, but for non-dilute case with the dry air mass of 62.10\,kg\,m$^{-2}$ and the absorbed stellar flux of 208.5\,$\mathrm{W\,m^{-2}}$,
and all other parameters kept the same as in the dilute simulation. 
}\label{fig:time_non}
\end{figure}

With an Earthlike noncondensible inventory and $\tau_0 = 1.2$,  the atmosphere does not become strongly nondilute until
the stellar flux is sufficiently high to cause a runaway, in which case the system does not reach equilibrium.  In order to 
illustrate strongly nondilute convection in equilibrium, one could increase $\tau_0$ with fixed stellar flux, or keep
$\tau_0$ at its original value while reducing the noncondensible inventory. Here we choose the latter approach, and raise the water vapor concentration in the 1D model by reducing the mass of the dry air to 62.10\,kg\,m$^{-2}$. In this case, the atmosphere can be non-dilute even at freezing point of water since the saturation vapor pressure of water in the atmosphere only depends on the air temperature. This provides a simple and clean test of energy and mass conservation in the nondilute case. With reduced
noncondensible pressure, the curve OLR$(T_s)$ shows very little overshoot; OLR$_\infty$ is the same as in the previous case, as
it is determined by the limiting case of a pure water vapor atmosphere. The absorbed stellar flux in this case was set at
208.5\,$\mathrm{W\,m^{-2}}$, which is just short of the value at which the system enters a runaway state. 

In this nondilute case, the system reaches both energy and mass equilibrium after integration of 3000~days (Figure~\ref{fig:time_non}a and \ref{fig:time_non}b). 
The net evaporation rate at the surface approaches zero exponentially with time, similar to the dilute case (Figure~\ref{fig:time_dilute}b).
However, the net radiative flux at the top of the model evolves somewhat more slowly compared with the dilute simulation (Figure~\ref{fig:time_dilute}a). This is related to the high climate sensitivity of water-rich atmospheres, i.e. the
low slope of OLR as a function of surface temperature.
 Figure~\ref{fig:time_non}c and \ref{fig:time_non}d show the equilibrium temperature and humidity profiles. 
The convection is so deep that it establishes a moist adiabat nearly throughout  the column atmosphere. 
The tropopause is close to  the top of the column model, $\sim 0.03$\,hPa. 
The humidity profile confirms that water vapor is non-dilute in the atmosphere. 
Corresponding to the thermal structure, the vertical distribution of the specific humidity becomes  more uniform compared with the dilute case. 
Even at the top of the model, the specific humidity is as high as 0.27\,kg/kg. In addition, the surface pressure in equilibrium is about 9.75\,hPa (Figure~\ref{fig:time_non}c and d), so that the mass of water in the atmosphere is $\sim$40\,kg, approximately two thirds of that of the dry air. The humidity near the ground is slightly in excess of the value associated with the saturated
moist adiabat, but this does not actually arise from supersaturation in the convection scheme.  Rather, the excess moisture
arises from the fact that the temperature profile is very slightly warmer than the moist adiabat, and that the saturated
specific humidity is very sensitive to temperature when the noncondensible inventory is so low. 

The vertical structure of condensation and convective heating for nondilute convection will be discussed in connection
with the runaway greenhouse simulation.

\subsection{Runaway greenhouse and seasonal cycle simulation} \label{subsec:runaway}

\begin{figure}[h]
  \centering
  \includegraphics[width=0.6\textwidth]{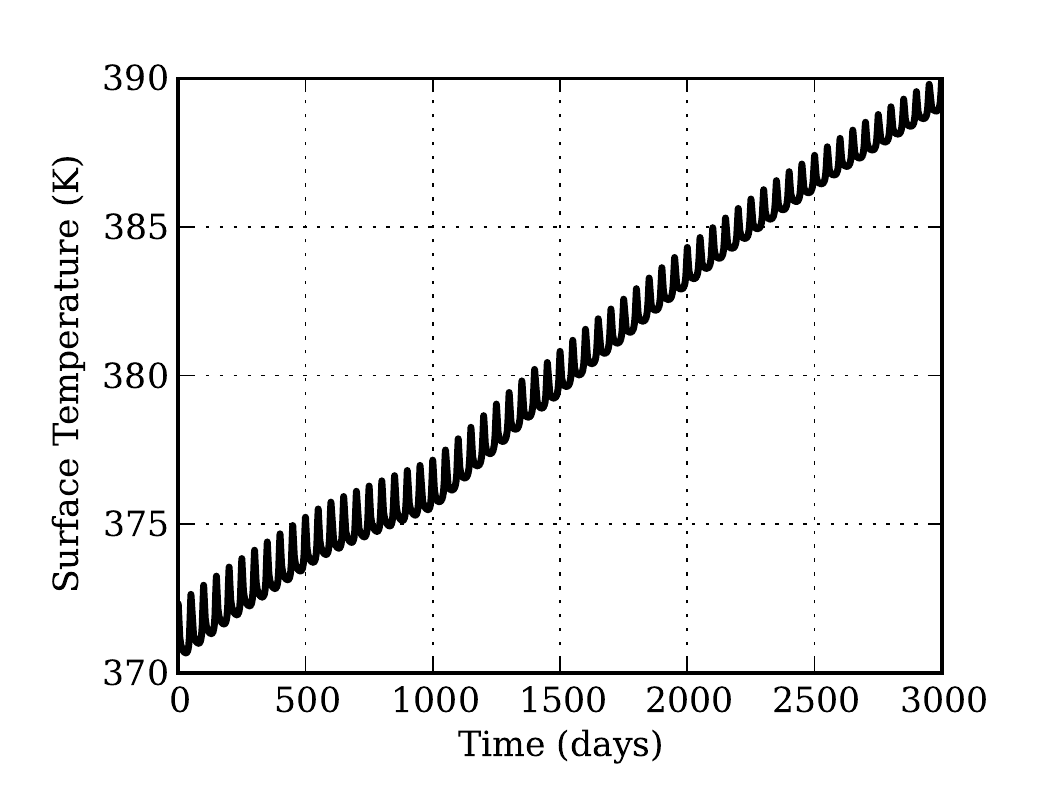}\\
  \caption{Time evolution of the surface temperature in the runaway greenhouse simulation. The seasonal cycle of the surface temperature is very weak ($\sim$1.5\,K) despite the highly eccentric orbit assumed.}\label{fig:time_runaway}
\end{figure}
 
To test how the moist convection scheme behaves for time-dependent simulations, we use the 1D column model to simulate the runaway greenhouse atmosphere on a planet with a high eccentricity ($e = 0.5$).
The orbital period of the planet in this simulation is 50\,days, and the stellar flux at the distance equal to the semi-major axis of the orbit is 238\,$\mathrm{W\,m^{-2}}$. Hence, the average of the stellar flux received by the planet over the course of the planetary year is $238 / \sqrt{1 - 0.5^2} = 274.82\,\mathrm{W\,m^{-2}}$, which is sufficiently high to force the system into a runaway.   Kepler's Second Law indicates that the planet travels slowly near the apastron (winter) and rapidly during periastron (summer), which would lead to short, hot summers and long, cold winters on a planet with little thermal inertia.

\begin{figure}[p]
  \centering
  \includegraphics[width=\textwidth]{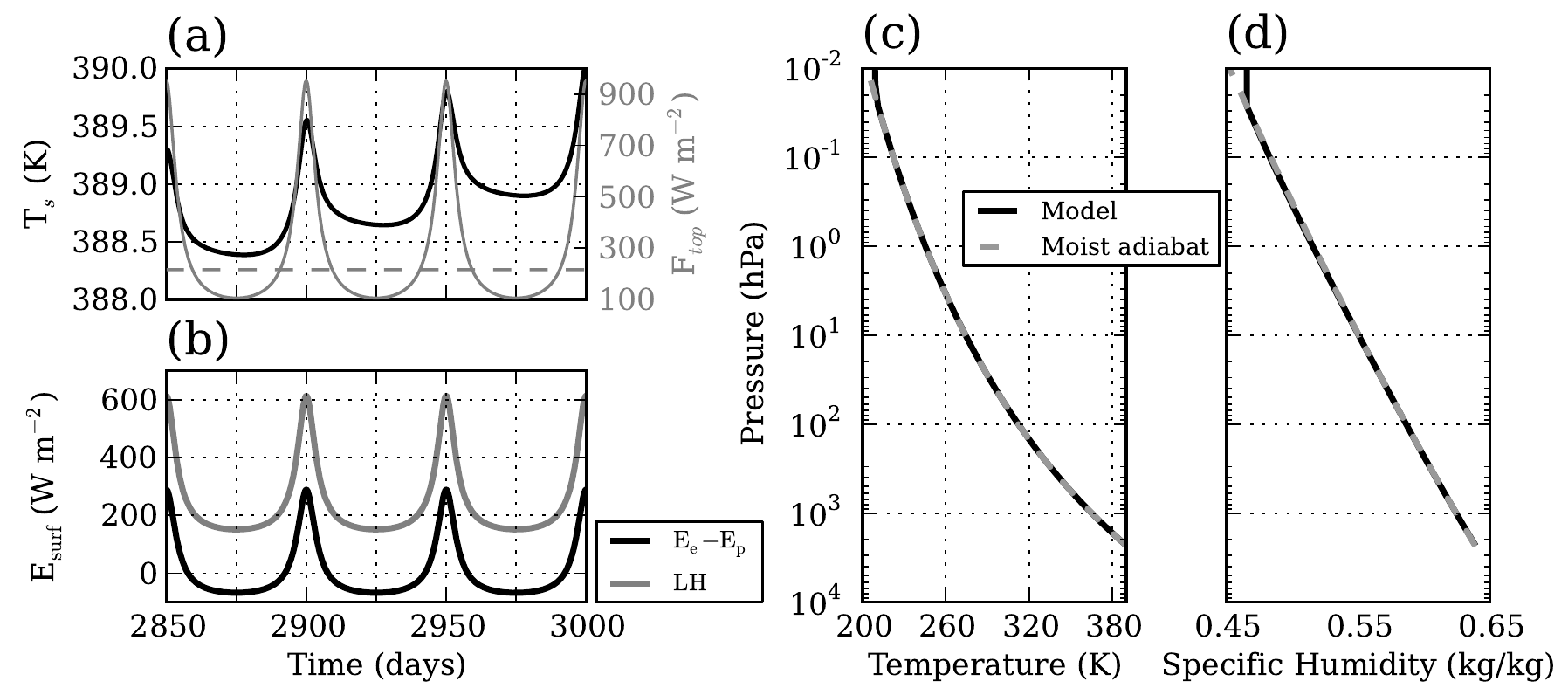}\\
  \caption{(a) Time evolution of the surface temperature (black), the global mean stellar flux received by the planet (gray) and the outgoing longwave radiation (dashed) for last 150~days of the runaway greenhouse simulation.  (b) Time evolution of the energy flux associated with mass exchange between the surface and the atmosphere ($E_\mathrm{e} - E_\mathrm{p}$, black) and  the latent heat flux due to surface evaporation (gray) for the same time period as in (a). (c) The vertical profile of air temperature  on day~3000. (d) The vertical profile of the specific humidity on day~3000.}\label{fig:runaway}
\end{figure}

We integrated the model for 3000~days starting from an initial surface temperature of 372\,K. The surface temperature evolution over
this time period is shown in Figure~\ref{fig:time_runaway}. The temperature exhibits a steady upward trend rising by $\sim 20\,$K over the course of the integration, with only a small superposed seasonal cycle. Because the annual mean OLR (Figure~\ref{fig:time_runaway}a) remains
near the limiting value of 210\,$\mathrm{W\,m^{-2}}$, which is less than the absorbed solar radiation, the planet is in a runaway
state and the temperature will continue to grow indefinitely so long as a liquid ocean remains to feed the increasing atmospheric
water content. Because of the weak seasonal cycle, the runaway conditions are little affected by the strong seasonal cycle
of insolation, even though the planet undergoes a long winter with stellar flux as low as 100\,$\mathrm{W\,m^{-2}}$ (gray line in Figure~\ref{fig:runaway}a).  If it were not for the evidently strong thermal inertia of the system much of the water vapor evaporated in summer would condense back onto the surface in the long winter.  

The detailed response of the surface temperature to the large seasonal stellar forcing is  shown in Figure~\ref{fig:runaway}a. The surface temperature varies by only 1.5\,K, in spite of  large stellar flux variation from 900\,W\,m$^{-2}$ to 100\,W\,m$^{-2}$ and small surface thermal inertia (recall that the thickness of the slab ocean is only 1\,m).  
The surface energy budget in the 1D model is
\begin{equation}
\frac{\upd ( \rho_w c_{pw} H  T_s)}{\upd t} = -SW - LW - SH - LH + (E_{p} - E_{e})
\end{equation}
Here, $\rho_w$ and $c_{pw}$ are the density and specific heat of liquid water respectively, $H = 1\,$m is the thickness of the slab ocean, $SW$ and $LW$ are the net shortwave and longwave radiative fluxes (positive when going upward), $SH$ and $LH$ are sensible and latent heat fluxes at the surface respectively, $E_{p}$ is the  energy loss of the atmosphere when the condensate falls to the ground
(equivalent to the sum of the internal energy and potential energy of the precipitated water, or the static energy of the precipitated water), and $E_{e}$ is the energy increase of the atmosphere when water goes into the atmosphere by evaporation (equivalent to the sum of the internal energy and potential energy of the evaporated water, or the static energy of the evaporated water). In the runaway greenhouse simulation, both $LW$ and $SH$ are small because the surface air temperature stays very close to the surface temperature and the atmosphere is optically thick for longwave radiation near the ground.
Hence, the absorbed stellar radiation is nearly balanced by the energy flux associated with the phase transition of water ($LH + E_{e} - E_{p}$, see Figure~\ref{fig:runaway}b). 
Therefore, most of the stellar flux is used to change the mass of the atmosphere instead of the surface temperature. The phase transition of water strongly damps temperature fluctuations, as it takes only a small change in surface temperature to drive a large change in the energy fluxes associated with water when the water content of the atmosphere is so high. Hence the planet exhibits very weak seasonal cycle in spite of its high eccentricity orbit. Note that $E_{e} - E_{p}$ is of comparable magnitude
to the latent heat flux, so that this term makes up a crucial part of the exchange of energy between surface and atmosphere
in the strongly nondilute case. 

Although the thermal inertia associated with phase change strongly damps the seasonal cycle, it is easy to verify from 
a simple energy balance argument that this thermal inertia does not cause a significant delay in transfer of water from the
ocean to the atmosphere.  For very hot conditions, the atmospheric energy storage is dominated by water vapor.  As an upper
bound, let's estimate the energy of a saturated atmosphere with surface temperature and pressure at the critical point of
water, beyond which point the distinction between ocean and atmosphere disappears.  The mass of the atmosphere per unit
area of planetary surface is $p_{crit}/g$, where $p_{crit}$ is the critical point pressure.  As an upper bound to latent
heat storage, we multiply by the latent heat $L$ at 1\,bar and 373\,K, which overestimates the latent heat storage because
in reality $L$ approaches zero as the critical point is approached (an effect not currently incorporated in our
convection scheme). Besides the latent heat storage, the atmosphere stores dry enthalpy at a rate $c_p T$ per unit mass, where
$T$ is the mass-weighted temperature of the atmosphere. As an upperbound to this, we take $T = T_{crit}$, which is not
a bad approximation given the slow logarithmic decay of $T(p)$ for a pure steam atmosphere.  The estimated energy storage per
unit area is then $E = (L+c_pT_{crit})p_{crit}/g$ which works out to $8\times 10^{12}\,\mathrm{J\,m^{-2}}$ for Earthlike gravity, 
with specific heat taken at the critical point.  The net flux available to allow this energy to accumulate is the
difference between the incoming stellar radiation and the OLR$_\infty$, which is 64\,$\mathrm{W\,m^{-2}}$.  Dividing this into
$E$, the time needed to reach the critical point is under 4000 Earth years, which is short compared to the other processes
involved in irreversible water loss. The time required to reach the critical point becomes infinite as the absorbed stellar flux
approaches OLR$_{crit}$, but even reducing the excess flux by a factor of 100 would not bring the thermal inertia delay 
into the range where it can be considered a significant inhibition to the runaway greenhouse process. 

The vertical structure of the temperature and specific humidity on the last day of simulation is shown in Figure~\ref{fig:runaway}c and \ref{fig:runaway}d. For the runaway greenhouse simulation, water vapor is non-dilute in the atmosphere. Similar to the non-dilute simulation in Section~\ref{subsec:non}, essentially the whole atmosphere is subject to convective adjustment. 
As the surface temperature reaches 390\,K, water vapor is non-dilute even in the upper atmosphere above 10\,hPa, leading to large infrared opacity there.
In contrast to the the dilute simulation discussed in Section~\ref{subsec:dilute} (Figure~\ref{fig:convec_dilute}),
condensation and convective heating are concentrated in a thin layer in the upper atmosphere, from 0.1\,hPa to 10\,hPa.
this is in part due simply to the optical thickness of the atmosphere,  which implies that the strong radiative cooling needed
to balance the latent heat release due to strong condensation can only occur in the high portions of the atmosphere where
the atmosphere first begins to become optically thin, and from which infrared can escape to space.  However, a full understanding
of the situation in the nondilute case requires some discussion of vertical motion and buoyancy generation, which exhibit
key differences from the dilute case, which will be discussed in the next section. 

\begin{figure}[h]
  \centering
  \includegraphics[width=\textwidth]{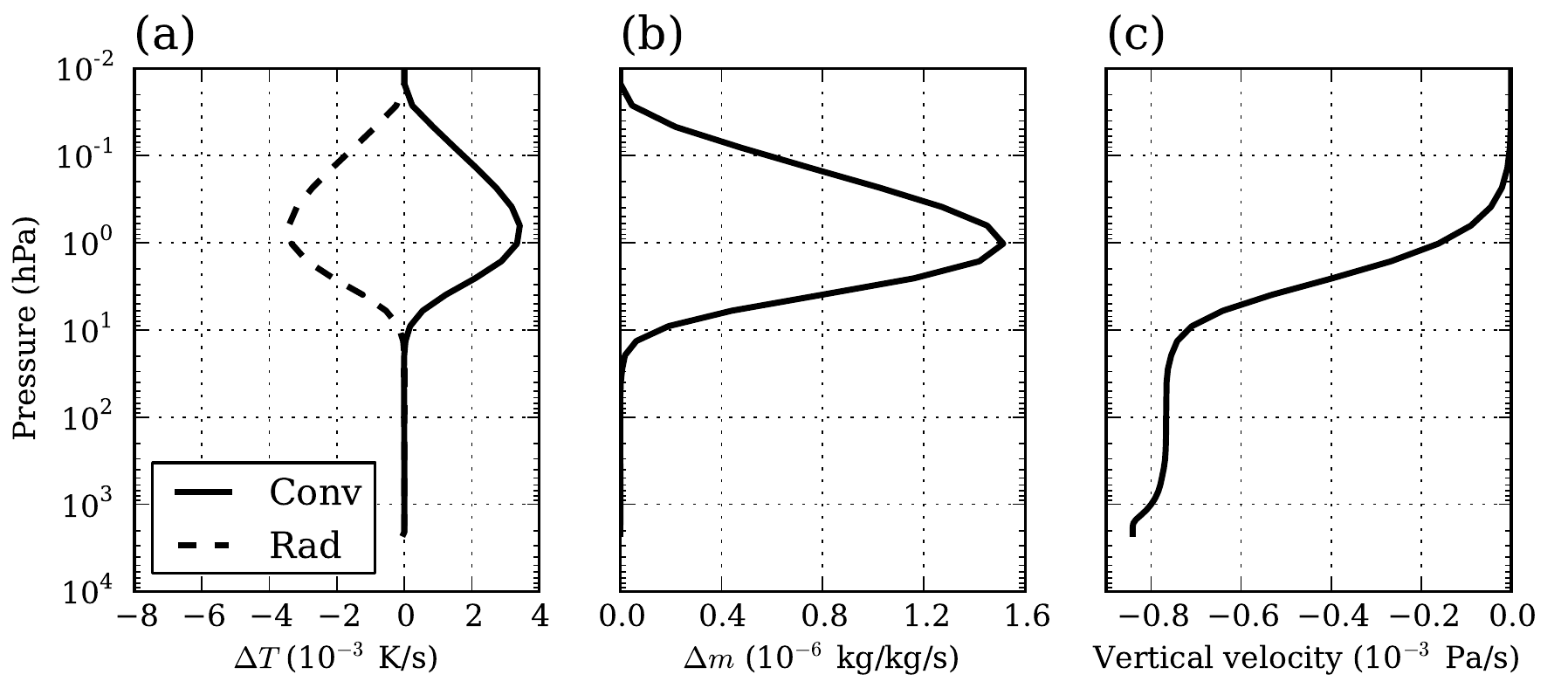}\\
  \caption{(a)  Heating rate due to convection (solid) and radiation (dashed). (b) Condensation rate of water. (c) Vertical velocity due to the removal of the condensed water from the atmosphere. All data is from the runaway greenhouse simulation on day~3000.}\label{fig:convec_runaway}
\end{figure}

\section{Discussion} \label{sec:discussion}

\subsection{Vertical motion induced by precipitation}

The fact that precipitation carries a significant mass flux has important implications for the vertical motion of the 
gas phase part of the atmosphere in the nondilute case. 
Vertical motion is diagnosed from the continuity equation. In the nondilute case, the continuity equation needs be modified taking the mass loss due to precipitation into account. Specifically, assuming the hydrostatic approximation, 
\begin{equation} \label{eq:12}
 \partial_p\omega + \nabla \cdot (\bm{v}) = P
\end{equation}
where $\bm v$ is the horizontal velocity and  $P$ is the rate of mass precipitation, per unit mass of the atmosphere. This is also the form the continuity equation would take when incorporating the convection scheme into a nondilute 3D general circulation model.  This effect is included in \citet{leconte13_nature} in their study of the runaway greenhouse threshold of the Earth's climate system. Integrating the continuity equation from the top of the atmosphere shows that a vertical motion is induced when the condensate is removed from the atmosphere ($\omega_2$ in Eq.(\ref{eq:omega})).
\begin{equation} \label{eq:omega}
\begin{aligned}
	\omega (p) &= \omega_1 + \omega_2 \\
     \omega_1  &\equiv - \int_{0}^{p} \nabla \cdot \bm{v}_p\ \upd p \\
     \omega_2  &\equiv \int_{0}^{p} P \ g
\end{aligned}
\end{equation}
Here $\omega_1$ comes from the air flow divergence above the pressure level, $\bm{v}_p = (u,v)$ is the horizontal wind velocity along the isobaric surface and vanishes in the 1D model simulation since there is no horizontal flow; this term would be present in 
a 3D model.  In the dilute limit there would be no mean vertical motion in an isolated 1D column, but in the nondilute case a mean gaseous vertical motion $\omega_2$ is supported by the downward mass flux of precipitation.  Figure~\ref{fig:convec_runaway}c shows the vertical velocity $\omega_2$ due to the removal of condensed water from the atmosphere for the runaway greenhouse simulation, and indicates that some condensation also occurs between the 1\,bar level and the surface. In this layer, the latent heat release  is balanced by the adiabatic cooling due to the  upward motion. For our single column model, the way that the vertical mass flux is balanced by a return mass flux in the form of precipitation can be considered as a peculiar form of one-column Hadley cell, in which the return mass flow compensating the upward motion in the convecting column is balanced by downward mass flux due to precipitation. For a conventional Hadley cell, the upward mass flux consists primarily of noncondensible gas, and must be balanced instead by downward noncondensible mass flux in adjoining subsiding regions, which leads to compressional heating there.   The magnitude of vertical motion in the optically thick nondilute case is small, however,  compared with the typical large-scale vertical velocity in the ITCZ of present Earth ($\sim$0.1\,Pa\,s$^{-1}$).  It is important to note that the transport of moisture and heat 
due to $\omega_2$ does not need to be incorporated in the column model or 3D general circulation model as an explicit vertical
advection term.  This transport is handled implicitly as part of the convective adjustment process. Specifically, it
is manifest as the change in pressure that occurs in a layer when mass is removed from a higher layer by precipitation. 

\begin{table}[h]
\centering
\begin{tabular}{lllll}
\hline
\hline
& dilute & non-dilute & runaway (summer) & runaway (mean) \\
\hline
Surface temperature (K) & 306.29 & 279.89 & 390.04  & 388.85 \\
$E_{p}-E_{e}$ (W$\,\mathrm{m^{-2}}$) & $-2.75$ & 2.91 & $-288.87$ & $-3.89$ \\
\hline
Precipitation rate (kg$\,\mathrm{m^{-2}\,s^{-1}}$) & 5.82$\times 10^{-5}$ & 6.12$\times 10^{-5}$ & $8.58\times 10^{-5}$ & 9.48$\times 10^{-5}$ \\
$\Delta F\mathrm{_{top}} - \Delta F\mathrm{_{bottom}}$ (W$\,\mathrm{m^{-2}}$) & 186.23 & 150.25 & 190.02 & 200.94 \\
\hline
$f$(water) (kg/kg) & 0.980 & 0.596 & 0.352 & 0.359\\
$q_{sa}$ (kg/kg) & 0.021  & 0.405 & 0.638 & 0.631 \\
\hline
\hline
\end{tabular}
\caption{Surface temperature, the energy flux associated with mass exchange between the atmosphere and surface ($E_{p}-E_{e}$), precipitation rate, the longwave radiative flux leaving the air column ($\Delta F\mathrm{_{top}} - \Delta F\mathrm{_{bottom}}$), $f$(water) and the specific humidity of the surface air ($q_{sa}$) for the dilute, non-dilute and runaway greenhouse simulations, respectively. For the runaway case, values are give both for the summer at day 3000 of the simulation, and for the average over the last seasonal cycle The ratio $f$(water) measures the portion of water vapor transported upward from the lowest layer by the moist convection relative to the total evaporated water added to the layer by the surface evaporation.}\label{tab:compare}
\end{table}

\subsection{Buoyancy generation in dilute vs. nondilute atmospheres}

In the dilute limit, absorption of stellar radiation at the surface heats the low-lying air, which then picks up moisture from
the adjacent surface; this builds buoyancy near the surface, leading to deep convection which takes the form of condensing
plumes penetrating deep into the atmosphere.  In the highly nondilute limit, in contrast, it is not possible to create buoyancy
in this way, because the saturated moist adiabat collapses onto a unique curve without free parameters -- the dewpoint/frostpoint
formula obtained by solving the Clausius Clapeyron relation for $T(p)$  -- which is neutrally stable with regard to pseudoadiabatic
vertical displacements.  In contrast, in the dilute limit, the lower atmosphere can be heated to a {\it different} member of
the moist adiabatic family of profiles, which is buoyant with regard to the overlying atmosphere.  In the strongly nondilute case,
heating the lower atmosphere increases surface pressure instead of buoyancy, in effect adding non-buoyant mass at the bottom
of the atmosphere. Similarly, it is difficult to generate top-driven convection through production of negative buoyancy
by radiative cooling in the upper atmosphere, because the energy loss goes into atmospheric mass loss via precipitation,
which reduces surface pressure rather than generating negative buoyancy aloft. (Retention of condensate would alter
this conclusion.)  The lack of buoyancy generation in single-component condensible atmospheres was noted in
 \cite{colaprete2003EarlyMarsClouds}  in connection with saturated pure-$\mathrm{CO_2}$ Martian convection, but is in fact a generic property of nondilute convection.  

\begin{figure}[h]
  \centering
  \includegraphics[width=\textwidth]{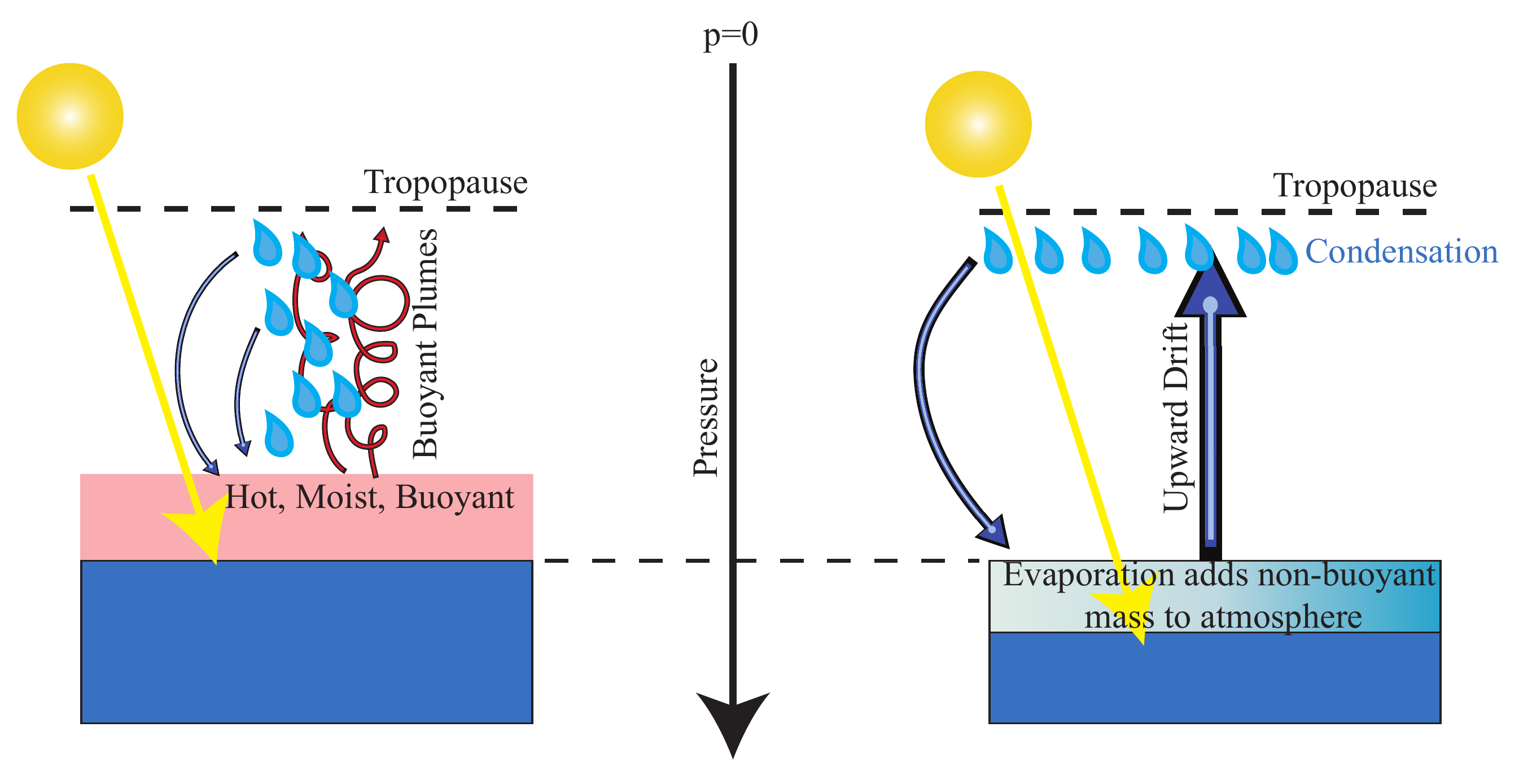}\\
  \caption{Schematic of moist convection in Earthlike conditions with a dilute atmosphere that is not optically thick throughout the infrared (left) and for a strongly nondilute optically thick atmosphere (right).  In the Earthlike case, shortwave stellar energy absorbed at the ocean surface leads to the creation of a hot moist layer of air near the surface, which is buoyant with respect to the overlying atmosphere and results in deep convection. In the strongly nondilute case, stellar heating instead adds mass to the bottom of the atmosphere in the form of condensible vapor. Because of constraints related to Clausius-Clapeyron, the new mass added is not buoyant with regard to the overlying atmosphere, and is only gradually carried upward by gradual laminar ascent, to upper layers of the troposphere where the latent heat released by rapid condensation can balanced by infrared cooling to space. }\label{fig:buoyancy}
\end{figure}

Our single-column simulations shed some further light on how nondilute moist convection works in the absence of buoyancy generation. 
Table~\ref{tab:compare} gives the ratio of the water vapor transported upward from the lowest layer during moist convection to the total evaporated water added to the layer during surface evaporation, which measures the buoyancy of the atmosphere during moist convection. This ratio is smaller than unity. The remaining part stays in the lowest layer, increasing the mass of the atmosphere, with negligible amounts condensing out there (see Figure~\ref{fig:convec_dilute}c and Figure~\ref{fig:convec_runaway}b). In our simulations, this ratio is nearly the same as  the mass concentration of non-condensable substance in the near surface layer (1-$q_{sa}$), as shown in the table.  This simple relationship 
is confirmed by numerical simulations with a variety of different surface air specific humidities ($0<q_{sa}<0.95$, not shown).
For the Earth-like atmosphere where water vapor is dilute, nearly all of evaporated water vapor is transported upward and forms precipitation in the mid-troposphere, releasing latent heat that balances the IR cooling there (Figure~\ref{fig:buoyancy}a). The column model indicates that the buoyancy of the atmosphere is getting weaker as water vapor becomes dominant in the atmosphere,
with less vertical transport by deep convective plumes that nearly instantly mix water vapor throughout the depth of the troposphere.  The evaporated water vapor tends to mostly stay in the lowest layer. 

In the runaway simulation, where the mass of the atmosphere is steadily increasing, most of the evaporated water does indeed
simply stay where it is put, at the bottom of the atmosphere.  In an equilibrium situation such as the cold nondilute
simulation with reduced noncondensible pressure,  the surface pressure eventually stops growing, and so the mass added to
the lowest model layer by evaporation must be carried away by some other means than deep convection.  The answer lies in
the advection due to the vertical velocity $\omega_2$, which carries moisture just a little ways upwards from the lowest
layer, rather than distributing it through the depth of the troposphere as deep convection would do.  The equilibrium mass budget
in strongly nondilute convection consists of addition of moisture to the bottom of the atmosphere by evaporation, which
is then advected upward to the layer where the mass can condense out and return to the surface as precipitation. 
From the standpoint of energetics, the reduction in surface pressure due to precipitation must balance the
increase due to evaporation because condensation is determined by infrared radiative cooling to space, while evaporation
is determined by absorption of stellar radiation at the surface, and the two energy fluxes must balance in equilibrium.
The contrast between dilute and nondilute moist convection is summarized in Figure \ref{fig:buoyancy}.  In the
nondilute case convection takes the form of a "moisture elevator" in which moisture added at the surface ascends the
floors of the atmosphere in an orderly and gradual process, in contrast to the chaotic, turbulent process by which
deep convection transports moisture in the dilute case. 
In the nondilute case, a mean vertical motion can exist in a single
column, with the upward vapor-phase mass flux balanced by downward mass flux from precipitation and the condensational heating is 
balanced locally by radiative and adiabatic cooling; this is another manifestation of the single-column
Hadley circulation  described previously. For the dilute case, in contrast, any mean upward motion in a column can still balance
condensational heating against adiabatic cooling by ascent locally, but the upward noncondensible mass flux must be compensated
by subsidence in the surrounding air, which leads to compressional heating that must be balanced by radiative cooling there.

\subsection{Energy transport by precipitation, and precipitation-temperature scaling}

The significant energy carried by precipitation is one of the novel features of nondilute convection. For present Earth's climate, even though the precipitation rate is trivial compared with the mass of the atmosphere, $E_{p}$ is a large value. For a typical value of precipitation rate in the tropics ($5\times 10^{-5}\,\mathrm{kg\,m^{-2}\,s^{-1}}$), $E_{p}$ is  as high as 50\,$\mathrm{W\,m^{-2}}$.  Table~\ref{tab:compare} shows ($E{_p} - E{_e}$) for the three simulations in Section~\ref{sec:result}. The column for the dilute simulation, which is carried out in Earthlike conditions,
shows that even though most of $E_{p}$ is canceled by the internal energy of the liquid water evaporated from the surface, the remaining part ($E{_p} - E{_e}$) is still not entirely negligible, attaining a value of approximately
$-2.75\,\mathrm{W\,m^{-2}}$. For the cold nondilute simulation with reduced noncondensible mass, the value gets 
slightly larger in magnitude and changes sign.  The change in sign arises primarily because the precipitation is warmer,
owing to the weaker vertical gradient of temperature in the nondilute case. Even in the hot runaway case, the seasonal mean value
only increases modestly in magnitude, to $-3.89\,\mathrm{W\,m^{-2}}$.  The modest values of $E{_p} - E{_e}$ trace back to the fact that the
precipitation rate is similar in all three cases,  with only moderate increases even in the hot runaway case. In equilibrium
or near-equilibrium, this limits the mass available to carry energy.  In the course of the seasonal cycle, however,
the exchange term can be very large, as evidenced by the summer runaway value in Table \ref{tab:compare}. In that case,
 $E{_p} - E{_e}$ is comparable to the absorbed solar flux at the surface. 
In our idealized calculation, all the energy carried by the precipitation is deposited at the ground, but in reality some
proportion would be transferred to the atmosphere by evaporation and frictional dissipation on the way down. 

The limited precipitation rate noted above arises from energetic limitations already familiar from studies of Earth's climate.
Because of limitations on turbulent transfer of sensible heat through a stable boundary layer, the net evaporation (and hence, in
equilibrium, net precipitation) cannot much exceed the stellar radiation reaching the surface \citep{pierrehumbert2002hydrologic,le2009snowball}. For the runaway case the limit is  $1.1\times 10^{-4}\,\mathrm{kg\,m^{-2}\,s^{-1}}$, only
slightly in excess of the realized precipitation.  Another way of looking at the energetics is that the dominant balance in
the troposphere when sufficiently large amounts of condensible substance is present is between latent heat release and net 
radiative cooling (which is purely infrared in our case). The column-integrated radiative cooling is shown in
Table \ref{tab:compare}. It accounts well for the precipitation in the two optically thick nondilute cases, translating
into a precipitation rate of $6.07\times 10^{-5}\,\mathrm{kg\,m^{-2}\,s^{-1}}$ in the cold nondilute case and $9.04\times 10^{-5}\,\mathrm{kg\,m^{-2}\,s^{-1}}$
in the runaway case. The radiative cooling limit significantly overestimates the precipitation in the dilute case, however,
because in that case much of the radiative cooling is balanced instead by sensible heat transport due to convection. 


\section{Conclusions} \label{sec:conclusions}

We have developed a simple energy-conserving convection parameterization which works across atmospheres ranging from those in which
the condensible substance is a dilute component of the atmosphere to those with strongly nondilute condensibles. 
The parameterization is
currently limited to a single condensible substance in a noncondensible background mixture of gases, but it can be applied to any combination
of the two classes of components, given suitable adjustment to thermodynamic parameters.   The performance of
the parameterization has been tested within a single-column radiative-convective model, but it is designed to be suitable for incorporation
in general circulation models. Simulations of that nature are underway, and will be reported on in a future paper.
Although our parameterization is not the first attempt to treat atmospheres with nondilute condensibles,
we have endeavored to use the formulation of the parameterization as a vehicle to give a thorough discussion of the nature of energy
conservation in nondilute atmospheres, and the various ways in which moist convection in such atmospheres differs from the more familiar
dilute limit. Key features include transport of significant amounts of energy and mass by precipitation (especially in the course of the seasonal
cycle when the atmosphere is out of equilibrium), the suppression of deep buoyant plumes in favor of a more gentle gradual ascent, 
and the feedback of precipitation and evaporation/sublimation on surface pressure (a feature already familiar from the current Martian
atmosphere). The single-column simulations were conceived primarily as a test of the conservation properties of the scheme, but in addition
to illustrating some general features of nondilute convection, demonstrated that nondilute atmospheres can have a very strong damping
effect on seasonal cycles driven even extreme seasonal variations in instellation.  

In principle, the large proportion of condensible substance present in nondilute atmospheres could lead the energy transport by condensate
to dominate the behavior of the system.  In all the cases we have examined, however, the energy carried by condensate, while a significant term needed to enforce energy conservation, is of fairly modest magnitude in the annual mean.  This property arises from energetic limits on precipitation rate:
in essence, one cannot release latent heat at a rate faster than it can be resupplied by absorption of stellar radiation, and so the large
latent heats of most condensible substances tend to yield limited mass of precipitation. This is a global constraint, which exerts dominant
control in a single-column model, but it would not apply locally in three-dimensional simulation, so the incorporation of energy transport
by condensate may have more dramatic effects when a scheme such as ours is embedded in a 3D general circulation model. 

The convection scheme we have developed adjusts the atmosphere to a saturated pseudoadiabat, using the lapse rate in comparison to
the nondilute pseudoadiabat as the sole criterion for convection.  This achieves an equivalent of what is commonly done in 
single-column radiative-convective models, and thus leads the way for general circulation model experiments focusing on dynamical
effects in nondilute atmospheres. When dealing with a situation presenting the novel physics of nondilute convection, it is useful
to revert to simplified schemes such as this which are easy to understand, and which can most easily be made to incorporate the most
fundamental physical constraints.  This approach complements approaches such as that pursued by \cite{wolf2015evolution}, in which
an attempt is made to adapt the complex Zhang-Macfarlane convection scheme -- which incorporates many empirical assumptions based on 
Earth's current atmosphere -- to nondilute conditions.  
However, unconditional convective adjustment to a pseudoadiabat does not necessarily represent the way convection operates in
reality. There are many potentially important physical effects that have not been taken into account. Among other things,
the derivation of the convection scheme has revealed a currently unconstrained parameter governing the distribution of
condensate following convection, and there is currently no good physical basis for setting this parameter.  In addition,
more work is needed on the actual behavior of the shallow nonprecipitating convection that occurs  when there is insufficient
condensible to allow the adjusted state to be saturated, and on schemes that in general allow the adjusted state to be subsaturated.
Further, given possibly strong contrasts between the molecular weight of the condensible and that of the background gas, there
is a need to incorporate compositional effects on buoyancy into the parameterization, along the lines explored for
$\mathrm{H_2 - H_2O}$ atmospheres by \cite{li2015moist}. Such effects would be particularly extreme for $\mathrm{H_2}$ atmospheres incorporating
condensible $\mathrm{CO_2}$ from a surface $\mathrm{CO_2}$ ocean or glacier, and would yield stable layers near the ground that are
highly resistant to the initiation of convection. Some insights may be gained by studying convection in nondiute atmospheres in the Solar
System -- notably those of Mars and Titan -- but it is likely that further development of nondilute convection parameterizations will need
to be informed by simulations with resolved three-dimensional convection/cloud simulations.

%
\section{Acknowledgments}
We thank Robin Wordsworth, Dorian Abbot, Daniel Koll and Jun Yang for many helpful discussions. Support for this work was provided by the NASA Astrobiology Institute’s Virtual Planetary Laboratory Lead Team, under the National Aeronautics and Space Administration solicitation NNH12ZDA002C and Cooperative Agreement Number NNA13AA93A.

\appendix

\section{Energy conservation for nondilute systems}
\label{apx:conservation}

The following closely follows the derivation given in \cite{trenberth1997AtmosBudgets}, but with 
close attention paid to retention of terms that are typically dropped in the dilute limit. 

We begin with the momentum and mass continuity equations, and the First Law written in 
the form given in Eq. \ref{eqn:FirstLawV}.  In altitude ($z$) coordinates these take
the form
\begin{align}
  \rho \frac{\upd}{\upd t}\bm{v} &= -  {\bm \nabla} p - \rho g {\hat {\bm z}}\label{eqn:Momentum}\\
  \partial_t \rho + \bm{\nabla\cdot} \rho {\bm v} &= 0 \label{eqn:MassContinuity}\\
  \frac{\upd}{\upd t} (k-\frac{p}{\rho}) + p \frac{\upd}{\upd t}\frac{1}{\rho} &= Q \label{eqn:Heat}
\end{align}
where ${\upd} / {\upd t}$ is the material derivative. The density $\rho$ includes both the gas phase and
condensed phase mass; mass may exchange between gas and condensed phase, but it is presumed in
this derivation that no mass leaves the air parcel in which it was originally found, and that the temperature
of the condensate is the same as the temperature of the surrounding gas. The effect of relaxing these assumptions
is discussed at the end of this Appendix.   $Q$ is the diabatic heating rate per unit mass. 

Taking the dot product of the momentum equation (Eq. \ref{eqn:Momentum}) with $\bm{v}$ and using the mass continuity equation (Eq. \ref{eqn:MassContinuity}) yields
\begin{equation}
\partial_t \frac{1}{2}\rho  \bm{ v \cdot  v} + \bm{\nabla\cdot}(\frac{1}{2}\rho \bm{ {v}\cdot{v}}){\bm v}
 =  -\bm{ {v}\cdot \nabla} p  - \rho g w 
\end{equation}
where $w$ is the vertical velocity.  Using $w = {\upd z}/{\upd t}$ together with the continuity equation allows
us to write the following expression for the evolution of kinetic plus potential energy
\begin{equation}\label{eqn:KEplusPE}
\partial_t (\frac{1}{2}\rho  \bm{ v \cdot  v} + \rho gz) + \bm{ \nabla\cdot}(\frac{1}{2}\rho  \bm{ v \cdot  v} + \rho g z){\bm v}
 =  -\bm{ {v}\cdot \nabla} p
\end{equation}
Using the mass continuity equation, the right hand side can be rewritten
\begin{equation}\label{eqn:BernoulliTerm}
-\bm{ {v}\cdot \nabla} p = -\bm{ \nabla \cdot} p {\bm v} + \frac{p}{\rho} \rho \bm{ \nabla\cdot{v}}
         =  -\bm{ \nabla \cdot} p {\bm v}- \frac{p}{\rho} \frac{\upd\rho}{\upd t}
\end{equation}
Now, upon multiplying by $\rho$ the First Law in Eq. \ref{eqn:Heat} can
be rewritten as
\begin{equation}
\partial_t \rho(k-\frac{p}{\rho}) + \bm{ \nabla\cdot} \rho(k-\frac{p}{\rho}) {\bm v}
  - \frac{p}{\rho} \frac{\upd}{\upd t} \rho = \rho Q\
\end{equation}
which can be combined with Eq. \ref{eqn:KEplusPE} and \ref{eqn:BernoulliTerm} to yield the conservation law
\begin{equation}
\partial_t \rho\mathcal{E} + \bm{ \nabla\cdot} \rho \bm {\mathcal{F}} = \rho Q 
\end{equation}
where the energy density is
\begin{equation}
\mathcal{E} \equiv \frac{1}{2} \bm{{v}\cdot{v}} + gz + (k-\frac{p}{\rho})
\end{equation}
and the energy flux is
\begin{equation}
\bm{\mathcal{F}} \equiv \left(\frac{1}{2}\bm{{v}\cdot{v}}+  g z + k \right)\bm{v}
\end{equation}

If the kinetic energy is negligible, the conservation law in the form Eq \ref{eqn:ColumnEnthalpyZ} follows upon integrating over
an isolated column of the atmosphere.  

The above calculation suffices to demonstrate moist enthalpy conservation in the context of the convective adjustment
protocol employed in the text, in which a layer is first mixed with condensate retained, followed by a separate step in
which condensate is removed from the column, taking its enthalpy and potential energy with it.  In a real atmosphere,
condensate would be redistributed amongst air parcels continuously, with some particles eventually reaching the bottom
where they are removed.  A formal treatment of conservation in this situation would be very complex, given that each
condensate particle has its own history-dependent velocity and each air parcel retains a mix of condensate particles
each with their own characteristics. However, a simple physical argument suffices to justify conservation in the general
case.  Imagine a situation in which a single condensate droplet moves from one air parcel to an adjoining one.  In that case,
it simply carries its own enthalpy with it, so the enthalpy removed from one air parcel is added to the other, and similarly,
potential and kinetic energy is moved, with some of the mechanical energy left behind in the source air parcel if there has
been drag against the particle, and some interchange between potential and kinetic energy if the particle has changed altitude.
Subsequent transformations of the condensate without leaving the destination parcel are within the scope of the preceding 
derivation. Since the energy transport for each droplet occurs independently of the others, if conservation applies for
displacement of an individual droplet, it will also apply for an ensemble of droplets.

\bibliography{runaway}
\bibliographystyle{apj}
%



\end{document}